\documentclass[twocolumn,showpacs,preprintnumbers,amsmath,amssymb]{revtex4}
\usepackage{amsmath}
\usepackage{natbib}
\usepackage{graphicx}% Include figure files
\usepackage{epstopdf}
\usepackage{subfigure}
\usepackage{dcolumn}% Align table columns on decimal point
\usepackage{bm}% bold math
\usepackage{color}
\begin{document}

\title{Integrability of Kerr-Newman spacetime with cloud strings, quintessence and
electromagnetic field}
\author{Wenfu Cao$^{1,2}$}
%\email{13477129050@163.com}
\author{Wenfang Liu$^{1}$}
%\email{21200007@sues.edu.cn}
\author{ Xin Wu$^{1,2,3}$}
\email{xinwu@gxu.edu.cn, wuxin_1134@sina.com}
 \affiliation{School of Mathematics, Physics and Statistics, Shanghai
University of Engineering Science, Shanghai 201620, China
\\ $^{2}$Center of Application and Research of Computational Physics,
Shanghai University of Engineering Science, Shanghai 201620, China
\\ $^{3}$Guangxi Key Laboratory for Relativistic Astrophysics, Guangxi
University, Nanning 530004, China}

\begin{abstract}

The dynamics of charged particles moving around a Kerr-Newman
black hole surrounded by cloud strings, quintessence and
electromagnetic field is integrable due to the presence of a
fourth constant of motion like the Carter constant. The fourth
motion constant and the axial-symmetry of the spacetime give a
chance to the existence of radial effective potentials with stable
circular orbits in two-dimensional planes, such as the equatorial
plane and other nonequatorial planes. They also give a possibility
of the presence of radial effective potentials with stable
spherical orbits in the three-dimensional space. The dynamical
parameters play important roles in changing the graphs of the
effective potentials. In addition, variations of these parameters
affect the presence or absence of stable circular orbits,
innermost stable circular orbits, stable spherical orbits and
marginally stable spherical orbits. They also affect the radii of
the stable circular or spherical orbits. It is numerically shown
that the stable circular orbits and innermost stable circular
orbits can exist not only in the equatorial plane but also in the
nonequatorial planes. Several stable spherical orbits and
marginally stable spherical orbits are numerically confirmed, too.
In particular, there are some stable spherical orbits and
marginally stable spherical orbits with vanishing angular momenta
for covering whole the range of the latitudinal coordinate.

\end{abstract}
%\keywords{Black hole; post-Newtonian approximations; three-body
% problem; dynamics of orbits}

%\pacs{95.30.Sf, 98.62.Sb, 97.60.Lf}

\maketitle

\section{Introduction}
\label{sec:intro}

Cosmological observations such as the cosmic microwave background
thermal anisotropies support the accelerating expansion of the
Universe [1]. The expansion is well explained by a dark energy
with repulsive gravitational effect. The dark energy is dependent
on the cosmological constant and the so-called quintessence
surrounding a black hole [2]. The cosmological constant acting as
vacuum energy with negative pressure causes the acceleration. The
quintessence as a scalar field coupled to gravity has negative
pressure. The quintessence plays a role in the cosmological
constant for an appropriate choice of the quintessence parameters
[2]. The Robertson-Walker metric with the accelerating scale
factor caused by the quintessence and the
Reissner-Nordstr\"{o}m-de Sitter black hole surrounded by the
quintessence can be found in [2,3]. The importance of
quintessential fields in the physical processes occurring around
black holes has been discussed in [4-7].

In addition to the quintessence, another extra source representing
the universe is not a collection of point particles, but is a
collection of extended objects, such as one-dimensional strings
considered by Letelier [8]. These extended objects like a cloud of
strings surrounding a black hole are helpful to describe physical
phenomena in the universe, and have astrophysical observable
consequences [9]. A static and spherically symmetric black hole
surrounded by the quintessence and the cloud of strings was given
in [10]. The Reissner-Nordstr\"{o}m metric with the quintessence
and the cloud of strings was also obtained in [11].

With the aid of the Newman-Janis algoritm [12], the
above-mentioned non-rotating black holes with the quintessence
and/or the cloud of strings can be transformed to rotating black
hole counterparts [13-16]. Adding  the cosmological constant, the
authors of [17] obtained  the Kerr-Newman-AdS solutions of the
Einstein-Maxwell equation in quintessence field. Toledo $\&$
Bezerra [18] studied the Kerr-Newman-AdS black hole with
quintessence and cloud of strings. In this case, the
electromagnetic potential is generated due to the charge in the
black hole.

If these extra perturbation sources like the quintessence, cloud
of strings and electromagnetic fields are not considered, the
Schwarzschild, Reissner-Nordstr\"{o}m metric, Kerr and Kerr-Newman
metrics are integrable. As far as the axially-symmetric Kerr
spacetime is concerned, it is integrable due to four constants of
motion, which are the particle (or photon) energy, angular
momentum and rest mass associated with the 4-velocity normalizing
condition and the Carter constant governing the motion of
geodesics in the latitudinal direction [19]. The existence of the
Carter constant as the fourth constant gives the Kerr spacetime
the possibility of non-planar orbits with constant coordinate
radii corresponding to spherical photon orbits [20,21] as well as
the existence of circular orbits in the equatorial plane. Wilkins
[22] first found the existence of unstable spherical photon orbits
around the Kerr black hole and studied many properties of the
spherical photon orbits. This result was extended to spherical
orbits of charged particles in a Kerr-Newman geometry by Johnston
and Ruffini [23]. Several numerical examples of spherical photon
orbits around a Kerr black hole were plotted by Teo [20,24]. Exact
formulas for spherical photon orbits around Kerr black holes were
given by Tavlayan and Tekin [25]. The observability of a series of
images produced by spherical photon orbits around near-extremal
Kerr black holes was shown by Igata et al. [26]. The authors of
[27] studied properties of spherical photon orbits in the Kerr
naked singularity spacetimes. Spherical photon orbits were
discussed in the field of Kerr-de Sitter black holes [28,29] and a
five-dimensional rotating black hole [30]. The ringdown and shadow
observables are relevant to a special set of unstable null orbits
with constant radii. These orbits are light rings [31-36] for the
spherically-symmetric Schwarzschild type black holes and spherical
photon orbits for the axially-symmetric Kerr type spacetimes. The
threshold spherical photon orbits mark a boundary between the
photons captured by the black hole and the photons escaping to
infinity. Therefore, there have been many other papers focusing on
these spherical photon orbits (see, e.g., [37-39]).

When the quintessence and the cloud of strings as two extra
perturbation sources are included, they do not destroy the
integrability of the considered spacetimes. However, the fourth
constant or the integrability becomes absent in most cases when
electromagnetic fields as an external perturbation source are
further included in these spacetimes. Even these external magnetic
fields  induce chaos of charged-particle motions under appropriate
circumstances [40-48]. In spite of this, not all the external
magnetic fields surrounding the black holes can eliminate the
existence of the fourth constant.  As Carter [19] claimed, not
only the geodesic equations of particles (or photons) around the
Kerr black hole but also the equations of charged-particle orbits
in the Kerr spacetime  with an electromagnetic field described by
a covariant vector potential are analytically solved. Their
solutions are expressed in terms of explicit quadratures. Although
such a covariant vector potential is replaced with a more
complicated form, the integrability of charged particle motions in
Kerr-Newmann spacetimes was shown by Hackmann and Xu [49]. In
other words, the fourth constant of motion is still existent.

Apart from the two Kerr type black holes with external magnetic
fields mentioned in [19,49], the dynamics of charged particles
moving around the Kerr-Newman black hole surrounded by cloud
strings, quintessence and electromagnetic field [18] is
integrable. Providing such an integrable example is the main
motivation of the present paper. Based on this integrability,
stable circular charged-particle orbits exist in two-dimensional
planes, which are not confined to the equatorial plane. Stable
spherical charged-particle orbits are also present. Unlike the
authors who studied the spherical photon orbits in the literature
[37-39], we mainly focus on the stable circular charged-particle
orbits in two-dimensional nonequatorial planes and the stable
spherical charged-particle orbits. They are important in an
astrophysical scenario. The structure of the thin accretion
Keplerian disks is governed by the stable equatorial circular
orbits of test particles [27]. Above all, the innermost stable
circular orbits act as the inner boundary of the Keplerian disks.
The threshold spherical charged-particle orbits are important to
model the capture or accretion of matter by the black hole.

The outline of the paper is organized as follows. In Section II,
we introduce the considered dynamical model. In Section III, we
analyze the integrability of this system, radial effective
potentials, circular orbits and spherical orbits of charged
particles. Finally, we conclude our conclusions in Section IV.

\section{Kerr-Newman black hole with extra perturbation sources}

The considered spacetime metric is introduced briefly. A
super-Hamiltonian for describing the motion of charged particles
around the Kerr-Newman black hole immersed in an external
electromagnetic field is given.

 \subsection{Description of spacetime metric}

A negative pressure from a gravitationally repulsive energy
component leads to the accelerated expansion of the universe. Its
origin may be due to quintessence dark energy surrounding a black
hole. In Boyer-Lindquist coordinates $(t,r,\theta,\phi)$, a
spherically-symmetric static Schwarzschild black hole surrounded
by the quintessence is expressed in [2] as
\begin{equation}
ds^{2} =g_{\alpha\beta}dx^{\alpha}dx^{\beta},
\end{equation}
where covariant metric $g_{\alpha\beta}$ has four nonzero
components:
\begin{eqnarray}
&& g_{tt}=-(1-\frac{2M}{r}-\frac{\alpha_q}{r^{3\omega_q+1}}),
 \\
&&
g_{rr}=(1-\frac{2M}{r}-\frac{\alpha_q}{r^{3\omega_q+1}})^{-1},  \\
&&
g_{\theta\theta}=r^2,   \\
&& g_{\phi\phi}=r^{2}\textrm{sin}^{2}\theta.
\end{eqnarray}
$M$ is the black hole mass. The state equation describing the
relation among the quintessential state parameter $\omega_{q}$,
the pressure $p_{quint}$ and the energy density $\rho_{quint}$ is
\begin{eqnarray}
p_{quint} &=& \omega_{q}\rho_{quint}, \\
\rho_{quint} &=&
-\frac{3}{2}\frac{\alpha_{q}\omega_{q}}{r^{3(\omega_{q}+1)}}.
\end{eqnarray}
$\alpha_{q}\neq0$ is a quintessence parameter, and $\alpha_{q}=0$
corresponds to the  Schwarzschild black hole. If $\rho_{quint}>0$,
then $\alpha_{q}\omega_{q}<0$. The parameter $\alpha_{q}$ is
positive for the quintessence field. Thus, the quintessential
state parameter $\omega_{q}$ is negative. The state parameter has
three cases [7,16]: $-1<\omega_{q}<-1/3$ for the quintessence,
$\omega_{q}<-1$ for the phantom energy, and $\omega_{q}=-1$ acting
as a cosmological constant. The quintessence corresponds to the
stress-energy tensors
\begin{eqnarray}
T^{t}_{t} &=& T^{r}_{r}=\rho_{quint}, \\
T^{\theta}_{\theta} &=&
T^{\phi}_{\phi}=-\frac{\rho_{quint}}{2}(3\omega_{q}+1).
\end{eqnarray}
The properties of quintessence in an astrophysical scenario have
been discussed in some literature [4-7].

Letelier [8] considered the Schwarzschild black hole surrounded by
another extra source, which is a spherically-symmetric cloud of
strings as a collection of extended objects instead of point
particles. In this case, the stress-energy tensors are
\begin{eqnarray}
T^{t}_{t} &=& T^{r}_{r}=\rho_{cloud}=\frac{b_c}{r^2}, \\
T^{\theta}_{\theta} &=& T^{\phi}_{\phi}=0,
\end{eqnarray}
where $\rho_{cloud}$ represents the energy density regarding the
string cloud and $b_c$ is a positive parameter measuring the
intensity of  the cloud of strings. Replacing the quintessence
term in Eqs. (2) and (3) with the string cloud intensity  $b_c$,
Letelier obtained two metric components of the Schwarzschild black
hole with the string cloud as follows:
\begin{eqnarray}
&& g_{tt}=-(1-\frac{2M}{r}-b_c),
 \\
&& g_{rr}=(1-\frac{2M}{r}-b_c)^{-1}.
\end{eqnarray}

When the quintessence and the cloud of strings as two extra
sources of energy surround the Schwarzschild black hole,  the
total stress-energy tensor is a linear combination of the
stress-energy tensors corresponding to the quintessence and the
one associated with the cloud of strings:
\begin{eqnarray}
T^{t}_{t} &=& T^{r}_{r}=\rho_{quint}+\rho_{cloud} \nonumber \\
&=&
\frac{b_c}{r^2}-\frac{3}{2}\frac{\alpha_{q}\omega_{q}}{r^{3(\omega_{q}+1)}},
 \\
T^{\theta}_{\theta} &=&
T^{\phi}_{\phi}=-\frac{\rho_{quint}}{2}(3\omega_{q}+1).
\end{eqnarray}
Based on Eqs. (2), (3), (12) and (13), two components of the
metric for the description of the Schwarzschild black hole
surrounded by the quintessence and the cloud of strings can be
written in [10] as follows:
\begin{eqnarray}
&& g_{tt}=-(1-b_c-\frac{2M}{r}-\frac{\alpha_q}{r^{3\omega_q+1}}),
 \\
&&
g_{rr}=(1-b_c-\frac{2M}{r}-\frac{\alpha_q}{r^{3\omega_q+1}})^{-1}.
\end{eqnarray}

Suppose the black hole has an electrical charge $Q$ inducing  an
electromagnetic field. The authors of [11] provided a metric for
the Reissner-Nordstr\"{o}m black hole surrounded by the
quintessence and the cloud of strings. The two metric components
$g_{tt}$ and $g_{rr}$ are
\begin{eqnarray}
&&
g_{tt}=-(1-b_c-\frac{2M}{r}+\frac{Q^2}{r^2}-\frac{\alpha_q}{r^{3\omega_q+1}}),
 \\
&&
g_{rr}=(1-b_c-\frac{2M}{r}+\frac{Q^2}{r^2}-\frac{\alpha_q}{r^{3\omega_q+1}})^{-1}.
\end{eqnarray}

In terms of the Newman-Janis algoritm [12], the
Reissner-Nordstr\"{o}m black hole metric with the quintessence and
the cloud of strings can be transformed into the Kerr-Newman black
hole metric in the quintessence and the cloud of strings [13-16].
Adding a cosmological constant $\Lambda$ [17], the authors of [18]
obtained a Kerr-Newman-AdS solution immersed in quintessence and
string cloud. The metric solution has six nonzero components [18]:
\begin{eqnarray}
&&
g_{tt}=\frac{1}{\Sigma\Xi^2}(\Delta_{\theta}a^2\sin^{2}\theta-\Delta_{r}),
 \\
&&
g_{t\phi}=\frac{a\sin^{2}\theta}{\Sigma\Xi^2}[\Delta_{r}-\Delta_{\theta}(r^2+a^2)]=g_{\phi
t},  \\
& &
g_{rr}=\frac{\Sigma}{\Delta_r},  \\
&&
g_{\theta\theta}=\frac{\Sigma}{\Delta_{\theta}},   \\
&&
g_{\phi\phi}=\frac{\sin^{2}\theta}{\Sigma\Xi^2}[\Delta_{\theta}(r^2+a^2)^2-\Delta_{r}a^2\sin^{2}\theta].
\end{eqnarray}
The above notations are specified by
\begin{eqnarray}
 \Sigma &=& r^2+a^2\cos^{2}\theta,
 \\
\Delta_{r}&=&(1-b_c)r^2+a^2+Q^2-2Mr \nonumber  \\
& & -\frac{\Lambda}{3}r^2(r^2+a^2)-\alpha_q r^{1-3\omega_{q}},  \\
\Delta_{\theta} &=& 1+\frac{\Lambda}{3}a^2\cos^{2}\theta,   \\
\Xi &=& 1+\frac{\Lambda}{3}a^2.
\end{eqnarray}
The angular momentum of the rotating black hole is given in the
range $a\in[-M,M]$. The black hole's electrical charge is given in
the range $Q\in[-M,M]$. The speed of light $c$ and the
gravitational constant $G$ take geometrized units, $c=G=1$.

\subsection{Super-Hamiltonian system}

The charge in the Kerr-Newman-AdS black hole generates an
electromagnetic potential [18]
\begin{equation}
A_{\mu} = -\frac{Qr}{\Sigma}\delta^{t}_{\mu}+\frac{Qra
\sin^{2}\theta}{\Sigma\Xi}\delta^{\phi}_{\mu}.
\end{equation}
The motion of a test particle with charge $q$ and mass $m$ around
the Kerr-Newman-AdS black hole surrounded with the quintessence,
string cloud and electromagnetic field is governed by the
super-Hamiltonian
\begin{eqnarray}
H &=& \frac{1}{2m}g^{\mu\nu}(p_{\mu}-qA_{\mu})(p_{\nu}
-qA_{\nu}) \nonumber \\
&=& \frac{1}{2m}g^{tt}(p_t-qA_{t})^2
+\frac{1}{2m}g^{\phi\phi}(p_{\phi}-qA_{\phi})^2 \nonumber \\
&& +\frac{1}{m}g^{t\phi}(p_t-qA_{t})(p_{\phi}-qA_{\phi})
+\frac{1}{2m}g^{rr}p^{2}_{r} \nonumber \\
&& +\frac{1}{2m}g^{\theta\theta}p^{2}_{\theta},
\end{eqnarray}
where the six non-zero covariant metric components (20)-(24)
correspond to their contravariant components
\begin{eqnarray}
&&
g^{tt}=\frac{\Xi^2}{\Sigma}[\frac{a^2}{\Delta_{\theta}}\sin^{2}\theta-\frac{(r^2+a^2)^2}{\Delta_{r}}],
 \\
&&
g^{t\phi}=\frac{a\Xi^2}{\Sigma}(\frac{1}{\Delta_{\theta}}-\frac{r^2+a^2}{\Delta_{r}})=g_{\phi
t},  \\
& &
g^{rr}=\frac{\Delta_r}{\Sigma},  \\
&&
g^{\theta\theta}=\frac{\Delta_{\theta}}{\Sigma},   \\
&& g^{\phi\phi}=\frac{\Xi^2}{\Sigma}(\frac{1}{\Delta_{\theta}
\sin^{2}\theta}-\frac{a^2}{\Delta_{r}}).
\end{eqnarray}

Considering a set of Hamiltonian canonical equations
$\dot{x}^{\mu}=\partial H/\partial
p_{\mu}=g^{\mu\nu}(p_{\nu}-qA_{\nu})/m$, we have the covariant
generalized momenta
\begin{equation}
p_{\mu} =m g_{\mu\nu}\dot{x}^{\nu}+qA_{\mu}.
\end{equation}
Because another set of Hamiltonian canonical equations satisfy
$\dot{p}_{t}=-\partial H/\partial t=0$ and
$\dot{p}_{\phi}=-\partial H/\partial \phi=0$, $p_{t}$ and
$p_{\phi}$ are two constants of motion. $p_{t}$ corresponds to an
energy of the particle, and $p_{\phi}$ is an angular momentum of
the particle. They are
\begin{eqnarray}
E &=& - p_{t}=-[m(g_{tt}\dot{t}+g_{t\phi}\dot{\phi})+qA_t], \\
L &=& p_{\phi}=m(g_{t\phi}\dot{t}
+g_{\phi\phi}\dot{\phi})+qA_{\phi}.
\end{eqnarray}

Dimensionless operations are given to Eq. (30). The distances,
coordinate time $t$ and $a$ take the black hole mass $M$ as units;
that is, $r\rightarrow rM$, $t\rightarrow tM$ and $a\rightarrow
aM$. The proper time $\tau$ also takes the mass unit,
$\tau\rightarrow \tau M$. In addition,  $Q\rightarrow QM$,
$\Lambda\rightarrow \Lambda/M^2$, $\alpha_{q}\rightarrow
\alpha_{q}M^{1+3\omega_{q}}$, $H\rightarrow Hm$, $E\rightarrow
Em$, $p_r\rightarrow p_rm$, $p_{\theta}\rightarrow p_{\theta}Mm$,
$L\rightarrow LMm$, and $q\rightarrow qm$. It is particularly
pointed out that  $E$ and $p_r$ are measured in terms of $m$, but
$p_{\theta}$ is measured in terms of $mM$. The particle's angular
momentum $L$ is also measured in terms of $mM$, whereas the black
hole's angular momentum $a$ is measured in terms of $M$. After the
dimensionless operations are implemented, $a\in[-1,1]$,
$Q\in[-1,1]$, and $-2Mr$ in Eq. (26) becomes $-2r$. The
Hamiltonian (30) becomes a dimensionless form
\begin{eqnarray}
H &=&
\frac{1}{2}\frac{\Xi^2}{\Sigma}[\frac{a^2}{\Delta_{\theta}}\sin^{2}\theta-\frac{(r^2+a^2)^2}{\Delta_{r}}](\frac{Q^{*}r}{\Sigma}-E)^2
 \nonumber \\
&& +\frac{1}{2}\frac{\Xi^2}{\Sigma}(\frac{1}{\Delta_{\theta}
\sin^{2}\theta}-\frac{a^2}{\Delta_{r}})(L-\frac{Q^{*}ar}{\Sigma\Xi}\sin^{2}\theta)^2 \nonumber \\
&&
+\frac{a\Xi^2}{\Sigma}(\frac{1}{\Delta_{\theta}}-\frac{r^2+a^2}{\Delta_{r}})(L-\frac{Q^{*}ar}{\Sigma\Xi}\sin^{2}\theta) \nonumber \\
&&
\cdot(\frac{Q^{*}r}{\Sigma}-E)+\frac{1}{2}\frac{\Delta_r}{\Sigma}p^{2}_{r}
+\frac{1}{2}\frac{\Delta_{\theta}}{\Sigma}p^{2}_{\theta},
\end{eqnarray}
where $Q^{*}=qQ$ is an electromagnetic field parameter.

The Hamiltonian (39) is a relatively complicated 4-dimensional
nonlinear system with two degrees of freedom $r$ and $\theta$. For
the time-like case, this Hamiltonian is always identical to a
given constant
\begin{equation}
H=-\frac{1}{2}.
\end{equation}
The existence of this constant is because the particle's
4-velocity $\dot{x}^{\mu}=(\dot{t},\dot{r},
\dot{\theta},\dot{\phi})=U^{\mu}=\partial H/\partial
p_{\mu}=g^{\mu\nu}(p_{\nu} -qA_{\nu}) $ satisfies the relation
$U^{\mu}U_{\mu}=-1$ or the particle's rest mass is conserved.

\section{Integrable dynamics of charged particles}

Firstly, we discuss the integrability of the Hamiltonian system
(39) with a vanishing cosmological constant by finding a fourth
integral of motion in this system. Secondly, physically allowed
motion regions are analyzed. Thirdly, radial effective potentials
in two-dimensional planes are focused on and some stable circular
orbits are given. Finally, radial effective potentials in the
three-dimensional space are considered and some stable spherical
orbits are obtained.

\subsection{Integrability of the system (39) without cosmological constant}

As is demonstrated above, the particle's energy $E$,  angular
momentum $L$ and rest mass are three constants of motion in the
system (39). Does a fourth constant exist? Yes, it does when
$\Lambda=0$ although the magnetic field, cloud strings and
quintessence field are included in the Kerr-Newman spacetime. In
what follows, we introduce how to find the fourth constant.

Clearly, $\Delta_{\theta}=\Xi=1$ in the case of $\Lambda=0$. The
system (39) satisfying Eq. (40) becomes
\begin{eqnarray}
-\Sigma &=&
[a^2\sin^{2}\theta-\frac{(r^2+a^2)^2}{\Delta_{r}}](\frac{Q^{*}r}{\Sigma}-E)^2
 \nonumber \\
&& +(\frac{1}{\sin^{2}\theta}-\frac{a^2}{\Delta_{r}})
(L-\frac{Q^{*}ar}{\Sigma}\sin^{2}\theta)^2 \nonumber \\
&&
+2a(1-\frac{r^2+a^2}{\Delta_{r}})(L-\frac{Q^{*}ar}{\Sigma}\sin^{2}\theta)
 \nonumber \\
&& \cdot(\frac{Q^{*}r}{\Sigma}-E)+\Delta_r p^2_r  +p^2_{\theta}.
\end{eqnarray}
This equation has a separable variable form
\begin{eqnarray}
&& \frac{1}{\Delta_{r}} [aL-E(r^2+a^2)+Q^{*}r]^2-r^2-\Delta_r
p^2_r \nonumber \\
&& = (Ea\sin\theta-\frac{L}{\sin\theta})^2+a^2\cos^{2}\theta+
p^2_{\theta}.
\end{eqnarray}
The left-hand side of this equality is a function of $r$, but the
right-hand side of this equality is another function of $\theta$.
In general, the equality is impossible. If and only if  both sides
are equal to a new constant denoted by $K$, the equality  (42) is
admissible. This means that Eq. (42) can be split into two
equations
\begin{eqnarray}
\frac{1}{\Delta_{r}} [aL-E(r^2+a^2)+Q^{*}r]^2-r^2-\Delta_r p^2_r
&=&K, \\ p^2_{\theta}+
(aE\sin\theta-\frac{L}{\sin\theta})^2+a^2\cos^{2}\theta &=&K.
\end{eqnarray}
They belong to a first integral of  motion similar to the Carter
constant [19]. Eq. (43) or Eq. (44) is the fourth constant of
motion in the system (39). In fact, the obtainment of the fourth
constant or Eq. (42) implicitly comes from the Hamilton-Jacobi
equation of the Hamiltonian (39).

The four independent constants are described by Eqs. (37), (38),
(40) and (43) (or (44)). They determine the integrability of the
system (39). If $\Lambda\neq0$, then no separable form (42)
exists, and Eqs. (43) and (44) do not exist, either. Thus, the
system (39) is nonintegrable. Only the case of $\Lambda=0$ is
considered in our later discussions.

\subsection{Physically allowed motion regions}

Eqs. (43) and (44) are respectively rewritten as
\begin{eqnarray}
\Sigma^2 \dot{r}^2 &=& [aL-E(r^2+a^2)+Q^{*}r]^2-(r^2+K)\Delta_{r}
\nonumber \\
&=& \Re (r). \\
\Sigma^2 \dot{\theta}^2 &=& K-a^2\cos^{2}\theta
- (aE\sin\theta-\frac{L}{\sin\theta})^2\nonumber \\
&=& K-a^2+2aEL+a^2(1-E^2)\sin^2\theta   \nonumber \\
&&-\frac{L^2}{\sin^2\theta}=\Theta (\theta).
\end{eqnarray}
The conditions for physically allowed motions are
\begin{eqnarray}
\Re (r) &\geq& 0 \\
\Theta (\theta) &\geq& 0.
\end{eqnarray}

Eq. (47) corresponds to $E\geq E^{+}$ or $E\leq E^{-}$, where
$E^{\pm}$ are given by
\begin{eqnarray}
E^{\pm} = (aL+Q^{*}r\pm\sqrt{(r^2+K)\Delta_{r}}~)/(r^2+a^2).
\end{eqnarray}
In fact, their expressions are based on $\Re (r)=0$. For $|E|<1$,
$r$ in Eq. (47) can be allowed in a finite range outside the event
horizon; that is, the corresponding orbit is bound [22]. For
$|E|\geq1$, $r$ in Eq. (47) can be allowed in a semi-infinite
range outside the event horizon; namely, the orbit is unbound.

The physically allowed ranges of parameters for Eq. (48) are given
according to several cases. Because $d\Theta/d\theta$ =
$2\cos\theta[a^2(1-E^2)\sin\theta+L^2/\sin^3\theta]$, the function
$\Theta$  has a maximum at $\theta=\pi/2$ for $|E|< 1$ (precisely
speaking, $E^{+}\leq E< 1$ or $-1< E\leq E^{-}$), i.e.,
$\Theta_{max}=K-(aE+L)^2$ with $K\geq (aE+L)^2$. In this case, we
have $\vartheta\leq\theta\leq \pi-\vartheta$, where $\vartheta$ is
a positive root of the equation $\Theta=0$ as follows:
\begin{eqnarray}
\vartheta &=&\arcsin\{(\sqrt{\varrho^2+4a^2(1-E^2)L^2}-\varrho)
\nonumber \\
&& /[2a^2(1-E^2)]\}^{\frac{1}{2}} ~(|E|< 1 ~\textrm{and} ~ a\neq0), \\
\vartheta &=&\arcsin (L/\sqrt{\varrho}) ~(a=0), \\
\varrho &=& K-a^2+2aEL>0. \nonumber
\end{eqnarray}

If $|E|\geq1$ (precisely speaking, $E\geq max\{E^{+}, 1\}$ or $E
\leq min\{E^{-},-1\}$), the function $\Theta$   reaches a maximum
at $\theta=\psi$ or $\theta=\pi-\psi$, where $\psi$ and the
maximum are expressed as
\begin{eqnarray}
\psi = \arcsin[\frac{L^2}{a^2(E^2-1)}]^{1/4}~(0\leq L\leq
a\sqrt{E^2-1}~),
\end{eqnarray}
\begin{eqnarray}
\Theta_{max} = K-a^2+2aL(E-\sqrt{E^2-1}).
\end{eqnarray}
The motions are confined to a region $\theta\in (0,\zeta]\cup
[\pi-\zeta,\pi)$, where
\begin{eqnarray}
\zeta &=&\arcsin\{(\sqrt{\varrho^2+4a^2(1-E^2)L^2}+\varrho)
\nonumber \\
&& /[2a^2(E^2-1)]\}^{\frac{1}{2}} ~(a\neq0, ~ |E|\neq 1).
\end{eqnarray}
Thus, the motions for these cases are always confined to the
region $\theta\in[\vartheta,\pi-\vartheta]$ or $\theta\in
(0,\zeta]\cup [\pi-\zeta,\pi)$.

In particular, the conditions for the orbits covering whole the
range of the latitudinal coordinate $\theta\in(0,\pi)$ (note that
0 and $\pi$ are coordinate singularities) and reaching the
symmetry axis at $\theta=0$ can be found from Eq. (48) or the
physically allowed ranges of $\theta$. The conditions for
$\theta\in (0,\pi)$ are one of the following two cases. (i) $L=0$,
$|E|< 1$ and $K\geq a^2\geq0$. (ii) $L=0$, $|E|\geq 1$ and $K\geq
a^2E^2\geq0$. It is clear that zero angular momentum $L=0$ is a
necessary condition for the orbits covering whole the range of the
latitudinal coordinate.

\subsection{Effective
potentials and stable circular orbits in two-dimensional planes}

Based on $\Theta=0$ with $\theta=\pi/2$, the fourth constant $K$
is given by
\begin{eqnarray}
K=a^2-2aL E^{\pm}-a^2[1-(E^{\pm})^2]+L^2.
\end{eqnarray}
In this case, $E^{\pm}$ are the standard radial effective
potentials at the equatorial plane $\theta=\pi/2$ in many
references (e.g., [31-36]). Solving Eqs. (49) and (55) (or Eq.
(42) with $\theta=\pi/2$), we have the energies
\begin{eqnarray}
V^{\pm}_{\pi/2} &=& \frac{-B\pm\sqrt{B^2-AC}}{A}, \\
A &=& (r^2+a^2)^2-a^2\Delta_{r}, \nonumber \\
B &=& aL\Delta_{r}-(aL+Q^{*}r)(r^2+a^2),  \nonumber \\
C &=& (aL+Q^{*}r)^2-(r^2+L^2)\Delta_{r}. \nonumber
\end{eqnarray}
The radial effective potentials $V^{\pm}_{\pi/2}$ at the
equatorial plane $\theta=\pi/2$ depend on separation $r$ and
parameters $a$, $L$, $Q$, $Q^{*}$, $b_c$, $\alpha_{q}$ and
$\omega_{q}$.

Besides the radial effective potentials in the equatorial plane,
they are present in other planes. The planes are determined by
$\Theta (\theta)=0$ corresponding to $\dot{\theta}=0$ in Eq. (46)
and can be described by $\theta=\sigma$. Here, $\sigma$ is a
parameter describing some plane. In this case, $K$ reads
\begin{eqnarray}
K= (aE\sin\sigma-\frac{L}{\sin\sigma})^2+a^2\cos^{2}\sigma.
\end{eqnarray}
The energies obtained from Eqs. (43) and (57) are expressed as
\begin{eqnarray}
V^{\pm}_{\sigma} &=& \frac{-B\pm\sqrt{B^2-A_{\sigma}C_{\sigma}}}{A_{\sigma}}, \\
A_{\sigma} &=& (r^2+a^2)^2-a^2\Delta_{r}\sin^{2}\sigma, \nonumber \\
C_{\sigma} &=& (aL+Q^{*}r)^2-(r^2+\frac{L^2}{\sin^{2}\sigma}
+a^2\cos^{2}\sigma)\Delta_{r}. \nonumber
\end{eqnarray}
Eq. (58) is the radial effective potentials in the plane
$\theta=\sigma$ and includes the results in Eq. (56). Such radial
effective potentials in the nonequatorial planes are seldom met in
the existing literature.

Without loss of generality, $V^{+}_{\sigma}$ is considered. The
local extrema of the effective potentials $V^{+}_{\sigma}$
correspond to circular orbits with constant radii $r$, which
satisfy the conditions
\begin{equation}
\Theta(\theta)=0, ~~~~ \frac{dV^{+}_{\sigma}}{dr} =0.
\end{equation}
The local minima of the effective potentials $V^{+}_{\sigma}$
represent stable circular orbits (SCOs), which  satisfy Eq. (59)
and the following condition
\begin{equation}
\frac{d^2V^{+}_{\sigma}}{dr^2} \geq 0.
\end{equation}
The equality symbol "=" indicates the innermost stable circular
orbits (ISCOs). The local maxima of the effective potentials
$V^{+}_{\sigma}$ mean unstable circular orbits, which  satisfy Eq.
(59) and
\begin{equation}
\frac{d^2V^{+}_{\sigma}}{dr^2} < 0.
\end{equation}

We focus on the radial motions of charged particles in the
quintessence field with $-1<\omega_{q}<-1/3$ and $\alpha_{q}>0$.
Since the effects of parameters $a$, $L$, $Q$ and $Q^*$ on the
charged particle dynamics have been discussed in some references
(e.g., [50,51]), the parameters $b_c$, $\alpha_{q}$, $\omega_{q}$
and $\sigma$ how to affect the radial effective potentials are
mainly considered in Fig. 1. The graph at the equatorial plane
$\sigma=\pi/2$ in Fig. 1(a) shifts to the observer at infinity as
the cloud strings parameter $b_c$ increases. In this case, the
energy decreases and the gravity from the black hole is weakened.
This fact can be explained simply and intuitively in terms of the
second term on line 1 of Eq. (39). The term
$\Psi=-(r^2+a^2)^2(Q^*r/\Sigma-E)^2/(2\Sigma\Delta_r)$ gives
gravitational effects to the charged particles. It is clear that
$\Delta_r$ is a decreasing function of $b_c$, and $\Psi$ is, too.
This implies that the gravity from the black hole becomes small as
$b_c$ increases. Therefore, $V^{+}$ is a decreasing function of
$b_c$. When the cloud strings parameter $b_c$ increases in Table
I, the graph going away the black hole leads to increasing the
radius of ISCO at the equatorial plane, but decreasing the radius
of SCO. Here, the ISCOs and SCOs are considered under the
condition $0<V^{+}<1$ as well as the condition (60). The energy
also decreases with an increase of the positive quintessence
parameter $\alpha_{q}$ in Fig. 1(b). However, the energy increases
with the negative quintessential state parameter $\omega_{q}$
increasing in Fig. 1(c). These results are because $\Delta_r$ is a
decreasing function of $\alpha_{q}$ ($>0$) but an increasing
function of $\omega_{q}$. The radii of ISCOs in Table I get larger
when $\alpha_{q}$ and $\omega_{q}$ increase. An increase of
$\omega_{q}$ enlarges  the radius of SCO, while that of
$\alpha_{q}$ diminishes the radius of SCO. Fig. 1(d) describes
that the shape of the effective potential depends on the plane
parameter $\sigma$. The potential decreases as $\sigma$ increases.
This result can be seen clearly from Eq. (58). Eq. (58) is
rewritten as
$V^+=-C_{\sigma}/(B+\sqrt{B^2-A_{\sigma}C_{\sigma}})$. Because
$C_{\sigma}$ is  an increasing function of $\sigma$ and
$A_{\sigma}$ is a decreasing function of $\sigma$, $V^+$ is a
decreasing function of $\sigma$. An increase of $\sigma$ results
in decreasing the radii of SCOs and ISCOs.

In short, the main results concluded from Fig. 1 and Table I are
given as follows. When anyone of the three parameters $b_c$,
$\alpha_{q}>0$  and $\sigma\in(0,\frac{\pi}{2}]$ increases, the
potential (or energy) decreases, whereas the potential increases
with  $\omega_{q}<0$ increasing. The radii of SCOs decrease as the
four parameters increase. The radii of of ISCOs gets larger with
$b_c$ and $\alpha_{q}>0$ increasing, but smaller with $\omega_{q}$
and $\sigma$ increasing.

Fig. 2 plots  three SCOs and ISCOs at the planes $\sigma=\pi/2$,
$\pi/4$ and $\pi/6$ for the other parameters considered in Fig.
1(d). Here, an eighth- and ninth-order Runge-Kutta-Fehlberg
integrator (RKF89) with adaptive step sizes is applied to solve
the canonical equations of the Hamiltonian (39). This integrator
can give  an order of $10^{-12}$ to the Hamiltonian error $\Delta
H=H+1/2$ when the integration time $\tau=10^{5}$. These orbits
still remain circular and stable in the three-dimensional
configuration with $x=r\sin\theta\cos\phi$,
$y=r\sin\theta\sin\phi$ and $z=r\cos\theta$ during the integration
time. Thus, the SCOs and ISCOs can exist not only in the
equatorial plane but also in the non-equatorial planes.

\subsection{Effective
potentials and stable spherical orbits in the three-dimensional
space}

If $K$ does not satisfy Eq. (57) with $p_{\theta}=0$ but is freely
given and satisfies Eq. (44) with $p_{\theta}\neq0$, $E^{\pm}$ in
Eq. (49) are radial effective potentials in the three-dimensional
space. The effective potentials are functions of separation $r$
and depend on parameters $a$, $L$, $K$, $Q$, $Q^{*}$, $b_c$,
$\alpha_{q}$ and $\omega_{q}$.

The local extrema of the effective potentials $E^{+}$ are
spherical orbits with constant radii $r$. The spherical orbits
should satisfy the condition
\begin{equation}
\frac{dE^{+}}{dr} =0,
\end{equation}
but do not always satisfy the condition $\Theta (\theta)=0$ for
any time. The spherical orbits must be unstable in the unbound
case of $E^{+}>1$. In the bound case of $0<E^{+}<1$, the spherical
orbits may either be stable or unstable. They are stable under
perturbations in the radial direction if
\begin{equation}
\frac{d^2E^{+}}{dr^2} \geq 0.
\end{equation}
Eqs. (49), (62) and (63) are the conditions  for the existence of
the stable spherical orbits (SSOs). The equality symbol ``=" in
Eq. (63) corresponds to the marginally stable spherical orbits
(MSSOs). In fact, the conditions for the SSOs are equivalent to
the following conditions
\begin{eqnarray}
\Re (r)=\frac{d\Re (r)}{dr}=0, ~~ \frac{d^2\Re (r)}{dr^2}\leq 0,
\end{eqnarray}
which were considered in the existing publications [20,24]. All
stable (or unstable) spherical orbits are of course confined to
the ranges $\theta\in[\vartheta,\pi-\vartheta]$ or $\theta\in
(0,\zeta]\cup [\pi-\zeta,\pi)$.  In particular, the conditions for
the spherical orbits covering whole the range of the latitudinal
coordinate are given in the above two cases. For $L\neq 0$, the
spherical orbits do not cover whole the range of the latitudinal
coordinate.

Fig. 3 describes the three-dimensional effective potentials
$E^{+}$, which correspond to the parameters for the
two-dimensional effective potentials $V^{+}$ in Fig. 1 but
$\sigma$ gives place to $K$. The dependence of the
three-dimensional potentials $E^{+}$ on each of the three
parameters $b_c$, $\alpha_{q}$ and $\omega_{q}$ is in agreement
with that of the two-dimensional potentials $V^{+}$. Unlike the
dependence of $V^{+}$ on $\sigma$ in Fig. 2(d), $E^{+}$ increases
with an increase of $K$ in Fig. 3(d). The reason is that $E^{+}$
in Eq. (49) is an increase function of $K$. The impacts of the
parameters $b_c$, $\alpha_{q}$ and $\omega_{q}$ on the radii of
SSOs and  MSSOs in Table II are similar to those of SCOs and ISCOs
in Table I. The radii of SSO and MSSO in Table II increase when
the parameter $K$ increases. Three spherical orbits and marginally
spherical orbits are shown in Fig. 4. When the integration time
reaches $\tau=10^{5}$, these spherical orbits still remain stable.

There are other notable points in Table II. The presence of
negative angular momenta $L$ for some of the MSSOs means that of
retrograde orbits moving against the black hole's rotation [23].
However, positive angular momenta $L$ correspond to prograde
orbits moving in the same direction as the black hole's rotation.
In addition to these positive and negative angular momenta,
vanishing angular momenta $L=0$ are also possible. For the case of
zero angular momenta, stable spherical orbits and marginally
stable spherical orbits cover whole the range of the latitudinal
coordinate and reach the symmetry axis at $\theta=0$, as shown in
Fig. 5. Such spherical orbits are called as the polar spherical
orbits [28]. However, the stable circular orbits are difficulty
present for vanishing angular momentum $L=0$. Why do the stable
spherical orbits exist in the case of $L=0$? Why do the stable
circular orbits not exist? The reason is that $K$ does not satisfy
Eq. (57) and is freely given for the stable spherical orbits, but
must satisfy Eq. (57) and is not freely given for the stable
circular orbits.

\section{Conclusions}

We analytically show the integrability of the dynamics of charged
particles moving around the Kerr-Newman black hole surrounded by
cloud strings, quintessence and electromagnetic field. This
integrability is due to the existence of a fourth constant of
motion like the Carter constant. If a nonvanishing cosmological
constant is included in the Kerr-Newman spacetime, then the fourth
constant is absent.

Because of the presence of the fourth motion constant and the
axial-symmetry of the spacetime, radial effective potentials and
stable circular orbits in two-dimensional planes involving the
equatorial plane and other nonequatorial planes can be present.
The dynamical parameters play important roles in changing the
graphs of the effective potentials. In addition, variations of
these parameters affect the presence or absence of stable circular
orbits and innermost stable circular orbits, and they also affect
the radii of the stable circular orbits and innermost stable
circular orbits. When each of the cloud strings parameter,
quintessence parameter and plane parameter increases, the graph of
effective potential shifts to the observer at infinity and the
effective potential decreases. However, the graph of effective
potential goes toward the black hole and the effective potential
increases as the quintessential state parameter increases. The
radii of stable circular orbits decrease. The radii of the
innermost stable circular orbits excluding those for the plane
parameter and the the quintessential state parameter increase. The
changes of these parameters exert more influences on those of the
radii of the stable circular orbits, but minor influences on those
of the radii of the innermost stable circular orbits. Numerical
tests show that the stable circular orbits and innermost stable
circular orbits can exist not only in the equatorial plane but
also in the non-equatorial planes.

On the other hand, the presence of the Carter-like constant and
the axial-symmetry of the spacetime also gives a chance to the
existence of radial effective potentials and stable spherical
orbits in the three-dimensional space. The three-dimensional
potential depending on each of the cloud strings parameter,
quintessential state parameter and quintessence parameter is
similar to the two-dimensional potential. The three-dimensional
potential increases with an increase of the fourth motion
constant.  The radii of stable spherical orbits and  marginally
stable spherical orbits varying with the cloud strings parameter,
quintessential state parameter and quintessence parameter is
consistent with those of stable circular orbits and innermost
stable circular orbits varying with these parameters. The radii of
stable spherical orbits and marginally stable spherical orbits
increase when the Carter-like constant increases. The existence of
some stable spherical orbits and marginally stable spherical
orbits are numerically confirmed. In particular, some stable
spherical orbits or marginally stable spherical orbits with
vanishing angular momenta for covering whole the range of the
latitudinal coordinate can also be found.

In sum, the Carter-like constant and the axial-symmetry of the
spacetime can ensure the presence of stable circular orbits in
two-dimensional nonequatorial planes and stable spherical orbits
in the three-dimensional space if a vanishing cosmological
constant appears in the Hamiltonian (30). Neither the stable
circular orbits in nonequatorial planes nor the stable spherical
orbits exist for a nonvanishing cosmological constant.

\acknowledgments

The authors are very grateful to the referee for valuable comments
and suggestions. This research was supported by the National
Natural Science Foundation of China (Grant No. 11973020) and the
Natural Science Foundation of Guangxi (Grant No. 2019JJD110006).

\begin{table*}[htbp]
\centering \caption{The radii $R_C$ of SCOs and the radii $R_I$ of
ISCOs for the parameters considered in Fig. 1. The notation ``-''
means the absence of SCOs and ISCOs. The energies $E$ of the SCOs
are not arbitrarily given but are determined by $R_C$. $E$ and the
angular momentum $L$ of the ISCOs are not arbitrarily given but
are determined by $R_I$. The values are not given for $E\geq1$.
The radii $R_C$ of SCOs decrease as anyone of the string cloud
$b_c$, quintessence parameter $\alpha_q$, quintessential state
parameter $\omega_q$ and plane parameter $\sigma$ increases. The
radii $R_I$ of ISCOs increase with the string cloud $b_c$ and
quintessence parameter $\alpha_q$ increasing, but decrease with
the quintessential state parameter $\omega_q$ and plane parameter
$\sigma$ increasing.
 \label{tab1}}
\begin{tabular}{ccccccccc}
\hline
   & Fig. 1(a) & $b_c$ & 0    & 0.05   &0.1     &0.15  &0.2 \\
   & SCO   & $R_C$ &10.15&8.26&-&-&-\\
   &   & $E$  &0.89&0.86&-&-&-\\
   & ISCO  & $R_I$ &6.33&6.72&7.17&7.67&8.25\\
   &  & $L$       &3.71&3.93&4.18&4.45&4.78\\
   &  & $E$       &0.88&0.86&0.83&0.8&0.78\\
\hline
   & Fig. 1(b) & $\alpha_q$       &0&0.1&0.2&0.3&0.4\\
   & SCO       & $R_C$     &12.13&9.84&-&-&-\\
   &           & $E$       &0.96&0.89&-&-&-\\
   & ISCO  & $R_I$ &5.67&6.4&7.37&8.72&10.77\\
   &   & $L$       &3.38&3.75&4.22&4.85&5.7\\
   &   & $E$       &0.93&0.88&0.82&0.75&0.68\\
\hline
   & Fig. 1(c) & $\omega_q$   &-0.35&-0.4&-0.45&-0.5&-0.55\\
   & SCO   & $R_C$     &11.94&12.14&12.65&14.06&-\\
   &   & $E$           &0.953&0.948&0.946&0.937&-\\
   & ISCO  & $R_I$ &5.73&5.76&5.8&5.88&6.06\\
   &   & $L$ &3.42&3.41&3.39&3.38&3.34\\
   &  & $E$ &0.94&0.93&0.92&0.91&0.90\\
\hline
 & Fig. 1(d) & $\sigma$    &$\pi/2$ & $\pi/3$ &$\pi/4$  & $\pi/5$&$\pi/6$\\
   & SCO   & $R_C$  &11.93&17.5&28.3&44.71&60.66\\
   &   & $E$        &0.95&0.96&0.97&0.98&0.99\\
   &ISCO            & $R_I$ &5.73&5.77&5.82&5.87&5.89\\
   &  & $L$         &3.41&2.97&2.62&2.03&1.73\\
   &  & $E$          &0.89&0.90&0.91&0.92&0.93\\
\hline
\end{tabular}
\end{table*}

\begin{table*}[htbp]
\centering \caption{The radii $R_S$ of SSOs and the radii $R_M$ of
MSSOs for the parameters considered in Figs. 3 and 5(a). The radii
$R_S$ of SSOs decrease when anyone of the string cloud $b_c$,
quintessence parameter $\alpha_q$, and quintessential state
parameter $\omega_q$ increases, but increase with the increase of
the Carter-like constant $K$. The radii $R_M$ of MSSOs increase as
the string cloud $b_c$, quintessence parameter $\alpha_q$ and
Carter-like constant $K$ increase, whereas decrease with the
increase of the quintessential state parameter $\omega_q$.
 \label{tab2}}
\begin{tabular}{ccccccccc}
\hline
   & Fig. 3(a) & $b_c$       & 0& 0.1&0.2&0.3&0.4\\
   & SSO & $R_S$           &19.45&15.79&-&-&-\\
   &   & $E$               &0.91&0.85&-&-&-\\
& MSSO  & $R_M$            &8.78&8.65&8.59&8.69&9.06\\
   &   & $L$               &-25.18&-13.37&0.64&18.16&41.76\\
   &   & $E$               &0.89&0.84&0.77&0.71&0.63\\
\hline
   & Fig. 3(b) & $\alpha_q$    &0&0.1&0.2&0.3&0.4\\
   & SSO   & $R_S$      &18.48&15.79&-&-&-\\
   &   & $E$            &0.92&0.85&-&-&-\\
& MSSO  & $R_M$ &8.62&8.65&8.76&9.06&9.78\\
   &  &  $L$    &-24.95&-13.37&0.63&18.62&43.93\\
   & & $E$      &0.91&0.84&0.77&0.69&0.60\\
\hline
   & Fig. 3(c) &$\omega_q$ &-0.35&-0.4&-0.45&-0.5&-0.55\\
   & SSO   & $R_S$  &15.79&19.11&-&-&-\\
   &   & $E$        &0.85&0.81&-&-&-\\
& MSSO  & $R_M$ &8.65&10.06&-&-&-\\
   &  & $L$&-13.37&-15.99&-&-&-\\
   &  & $E$&0.84&0.80&-&-&-\\
\hline
& Fig. 3(d) & $K$   &8    &12   &16   &20 &24\\
   & SSO   & $R_S$  &-&-&-&11.85&15.79\\
   && $E$           &-&-&-&0.84&0.85\\

   & MSSO  & $R_M$  &5.31&6.15&7&7.83&8.65\\
   &   & $L$ &22.8&14.74&5.93&-3.47&-13.37\\
   &   & $E$ &0.81&0.82&0.83&0.84&0.85\\
\hline
   & Fig. 5(a) & $K$   &8&12&16&20&24\\
   & SSO   & $R_S$  &-&-&-&10.39&14.68\\
   &   & $E$   &-&-&-&0.84&0.85\\
   & MSSO  & $R_M$ &5.31&6.15&7&7.83&8.65\\
   &   & $L$ &22.8&14.74&5.93&-3.47&-13.37\\
   & & $E$ &0.81&0.82&0.83&0.84&0.85\\
\hline
\end{tabular}
\end{table*}

\begin{figure*}[htbp]
\center{
\includegraphics[scale=0.31]{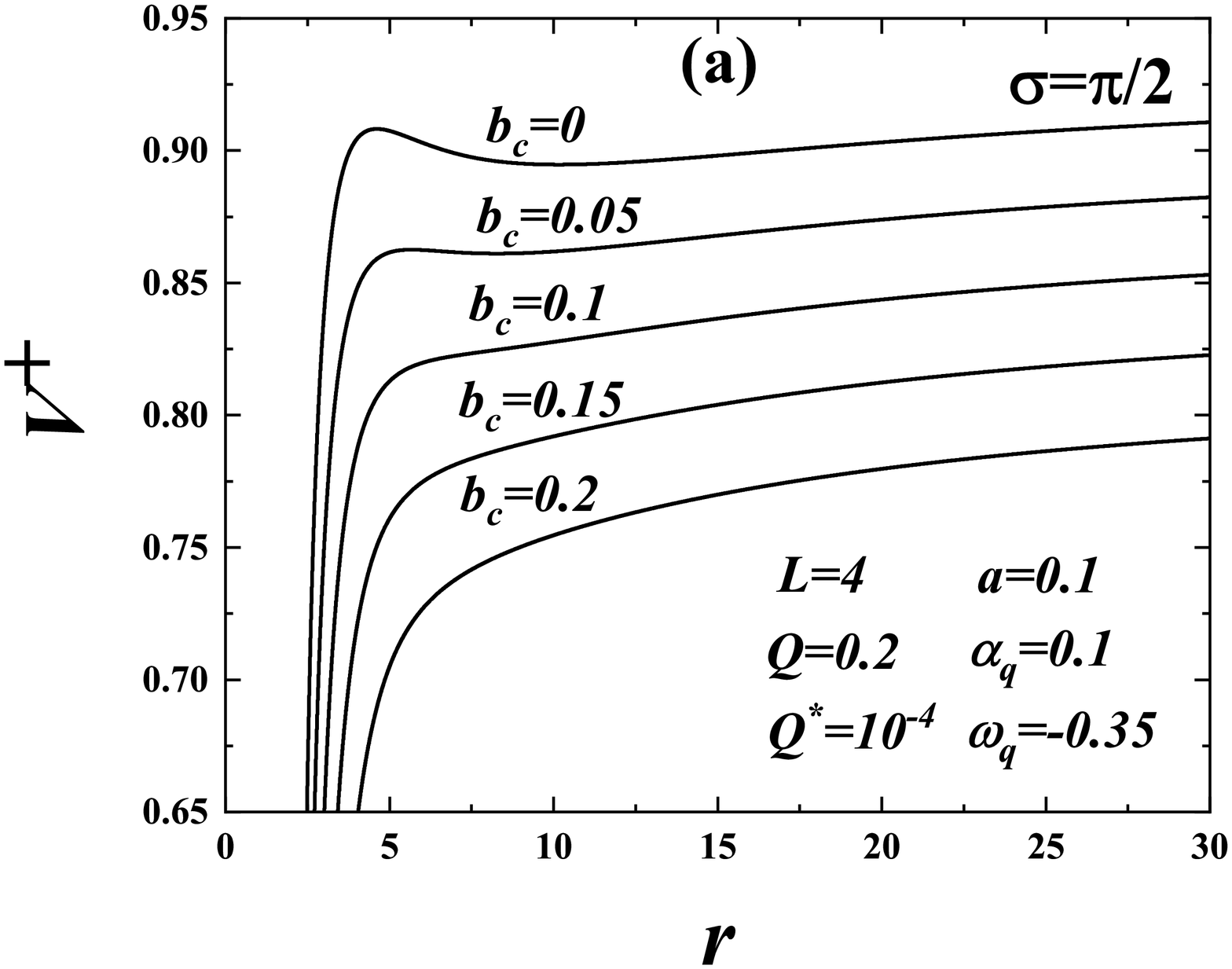}
\includegraphics[scale=0.31]{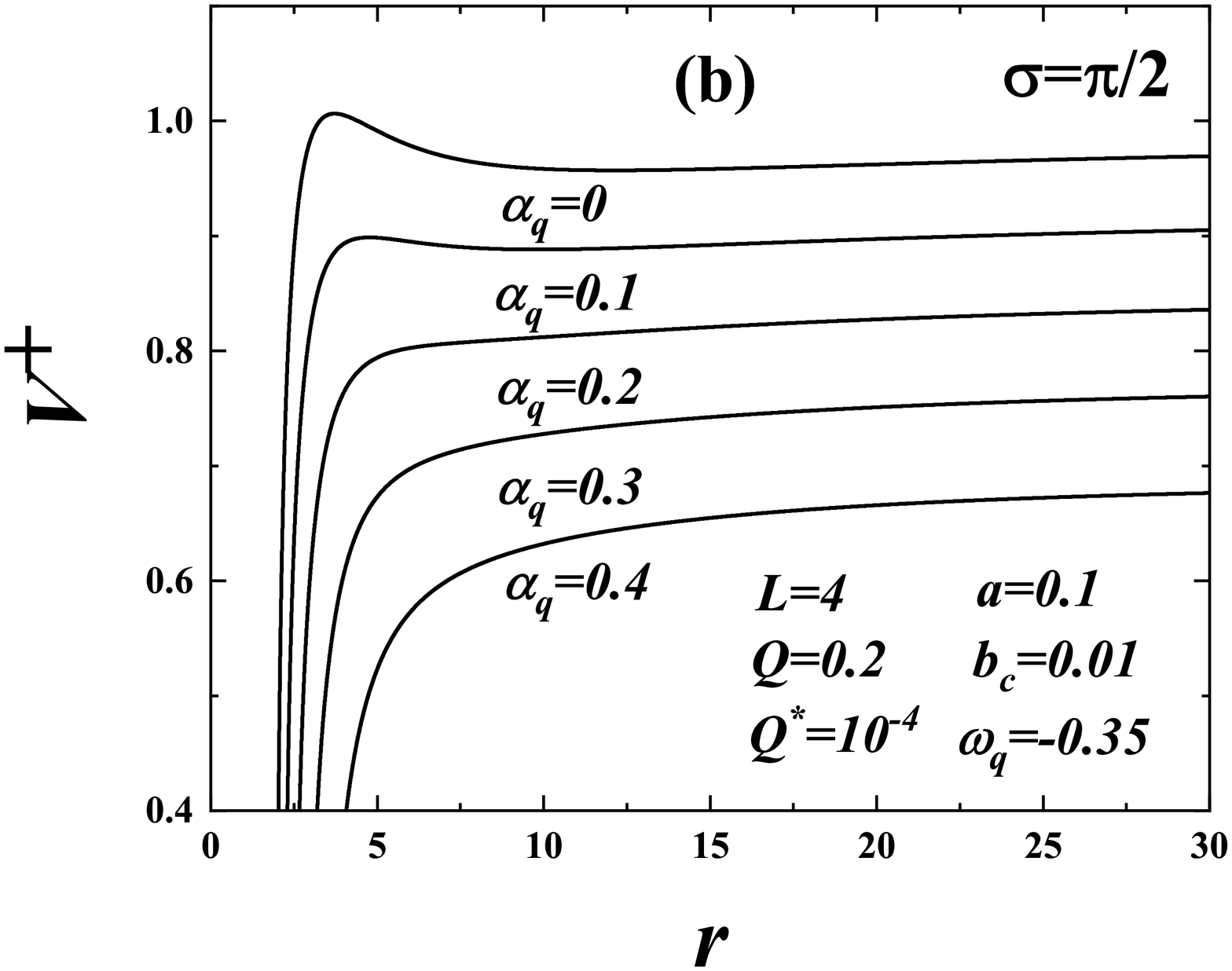}
\includegraphics[scale=0.31]{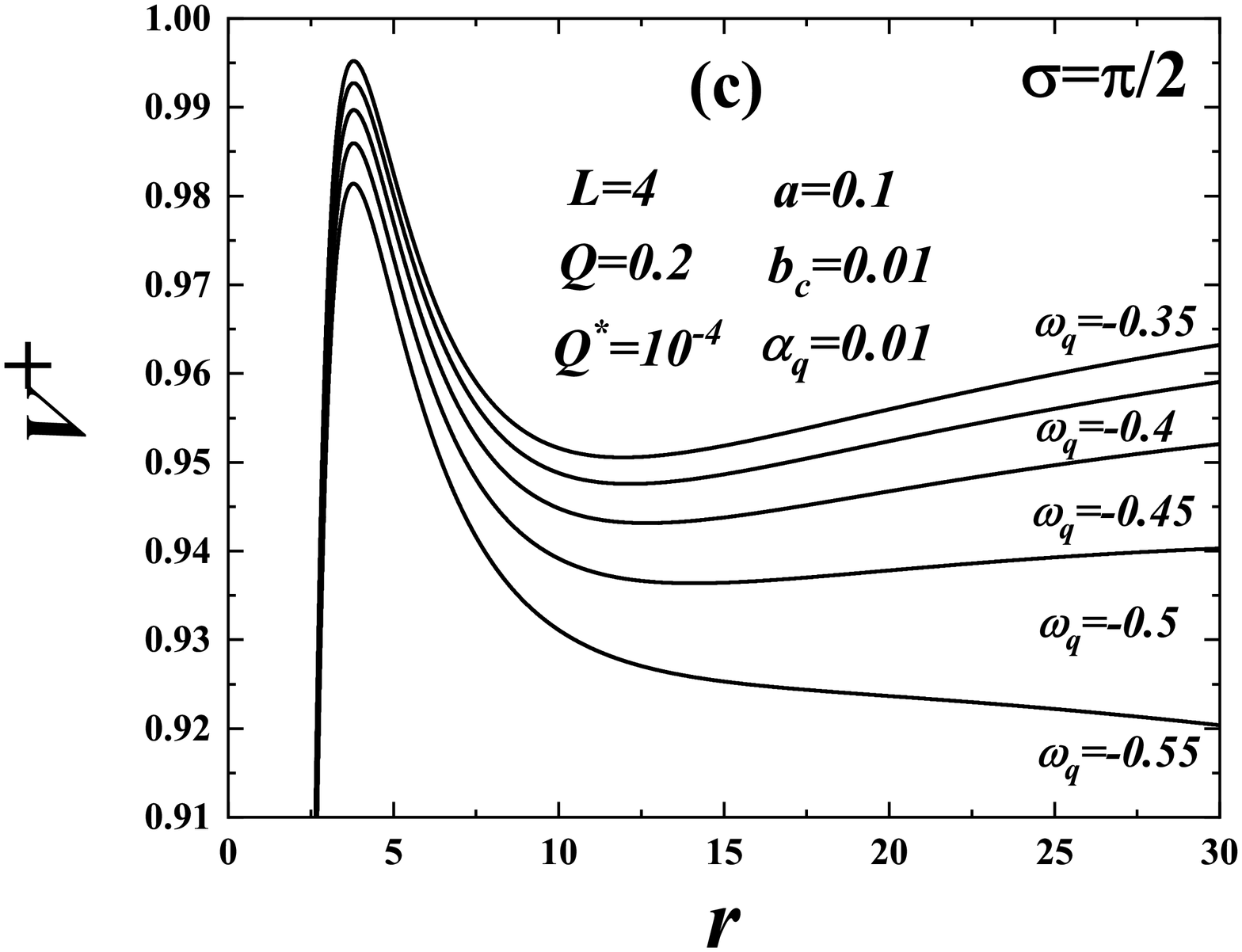}
\includegraphics[scale=0.31]{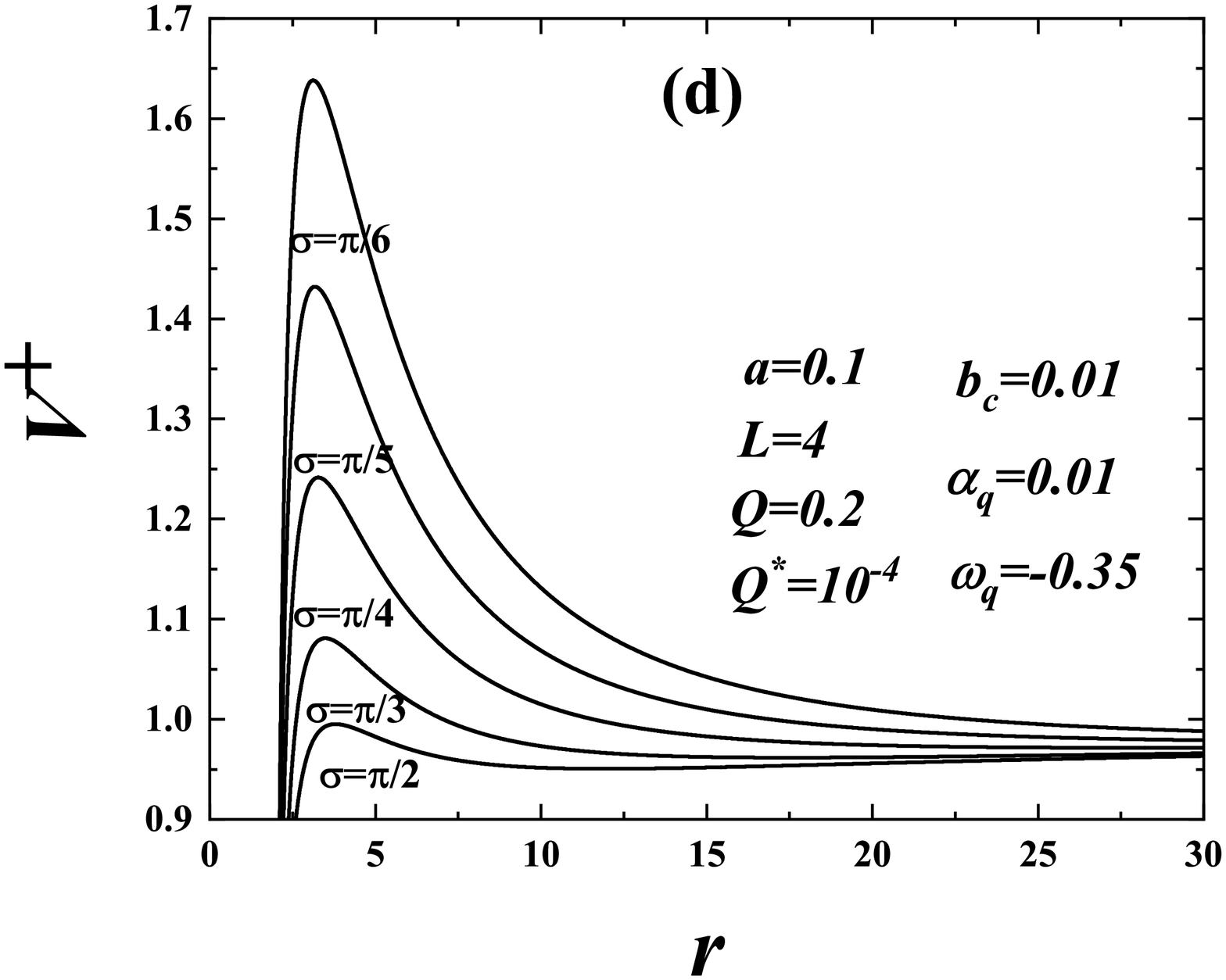}
\caption{Radial effective potentials $V^{+}$ of Eq. (58) in
two-dimensional planes $\theta=\sigma$. (a)-(c): They are plotted
at the equatorial plane $\sigma=\pi/2$. (d): They are plotted in
five planes $\sigma=\pi/6$, $\pi/5$, $\pi/4$, $\pi/3$ and $\pi/2$.
The impacts of the cloud strings parameter $b_c$ in Eq. (10),
quintessential state parameter $\omega_q$ in Eq. (6), quintessence
parameter $\alpha_q$ in Eq. (7) and plane parameter $\sigma$ on
the effective potentials are shown in panels (a)-(d),
respectively. The plane parameters $\sigma$ satisfy Eq. (57) with
$p_{\theta}=0$ and are constants. For a given separation $r$, the
potentials (i.e., energies) decrease as each of the string cloud
$b_c$, quintessence parameter $\alpha_q$ and plane parameter
$\sigma$ increases, whereas increase when the quintessential state
parameter $\omega_q$ increases.}
 \label{Fig1}}
\end{figure*}

\begin{figure*}[htbp]
\center{
\includegraphics[scale=0.2]{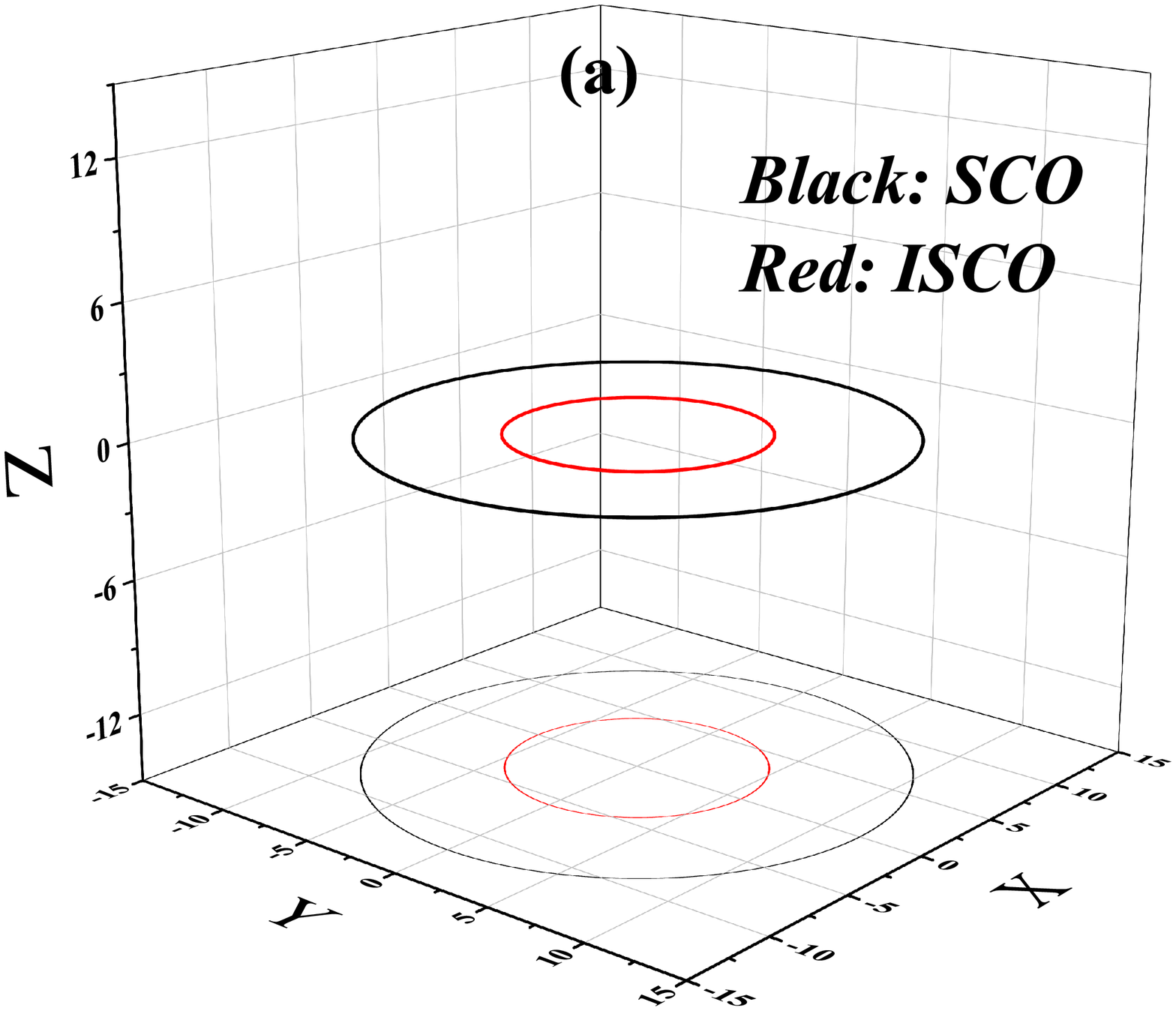}
\includegraphics[scale=0.2]{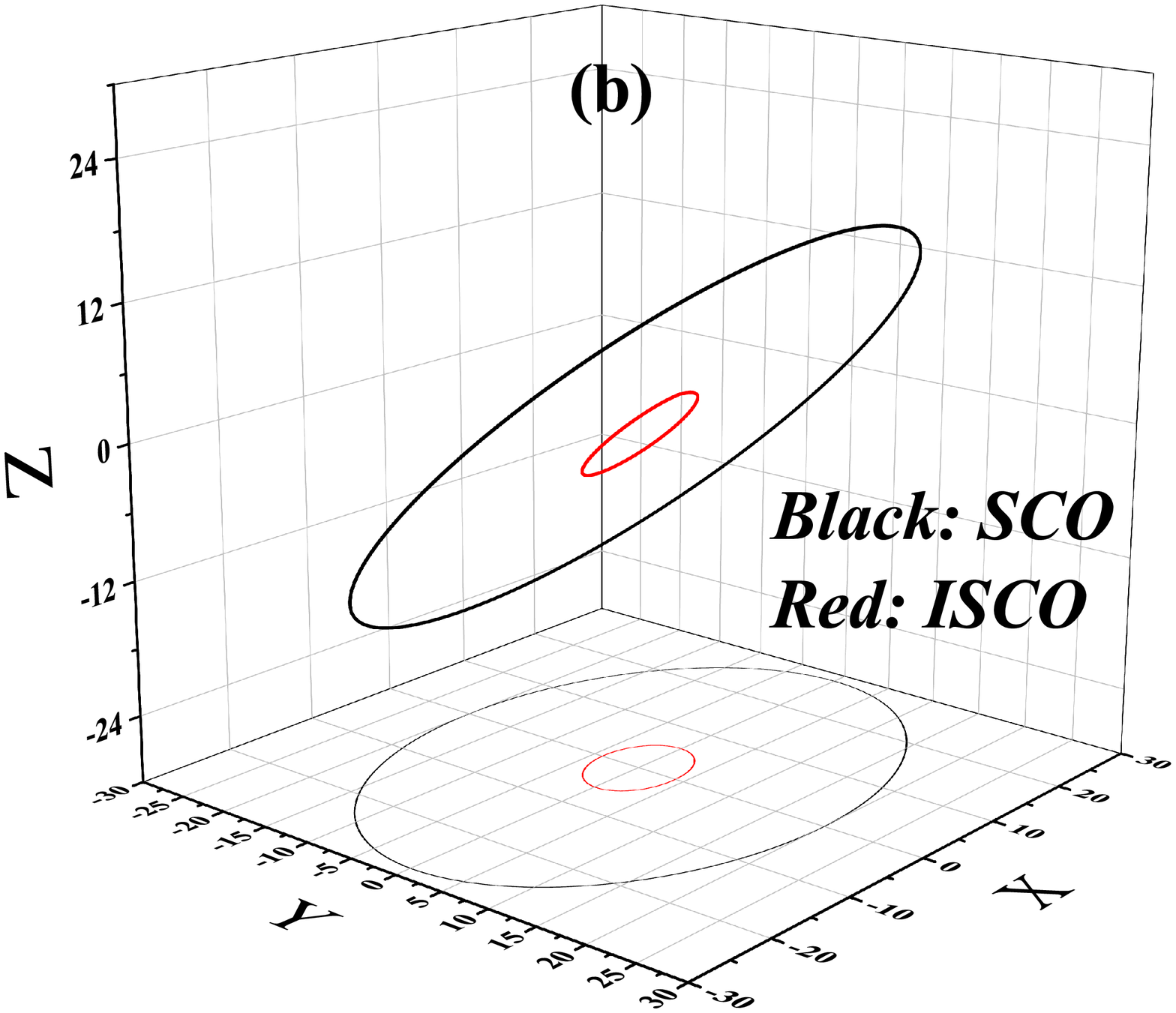}
\includegraphics[scale=0.2]{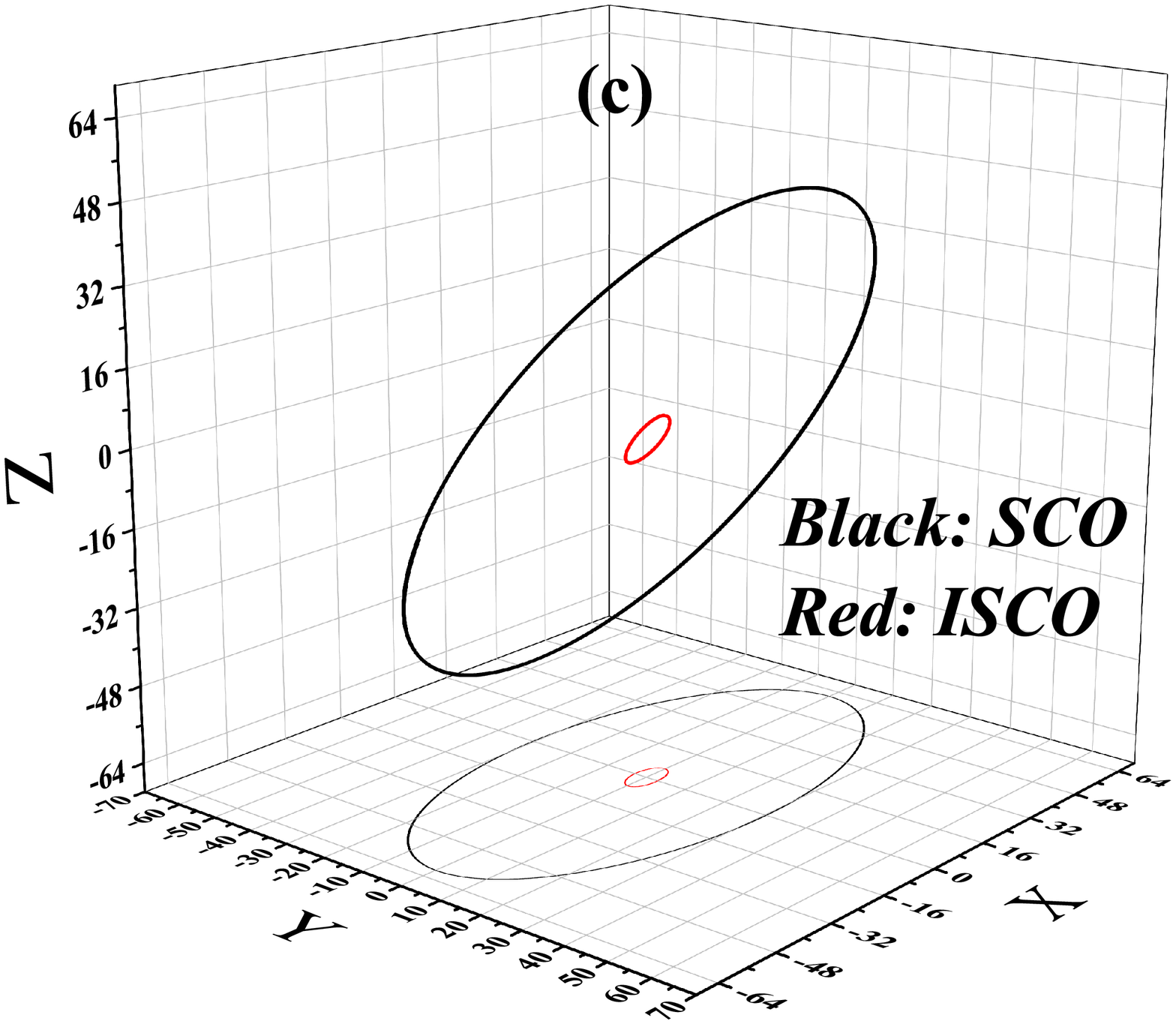}
\caption{Three examples of SCOs and  ISCOs  in two-dimensional
planes. (a): The SCO colored Black on the plane $\sigma=\pi/2$ has
its radius $r=11.93$ and the ISCO colored Red has the radius
$r=5.73$ and angular momentum $L=3.41$. (b): The SCO colored Black
on the plane $\sigma=\pi/4$ has its radius $r=28.3$ and the ISCO
colored Red has the radius $r=5.82$ and angular momentum $L=2.62$.
(c): The SCO colored Black on the plane $\sigma=\pi/6$ has its
radius $r=60.66$ and the ISCO colored Red has the radius $r=5.89$
and angular momentum $L=1.73$. The plane parameters $\sigma$
satisfy Eq. (57) with $p_{\theta}=0$ and are constants. The other
parameters in each of the panels are the same as those of Fig.
1(d). The upper part of each panel corresponds to the practical
trajectories, and the bottom part relates to projections of the
practical trajectories. These orbits still remain circular and
stable in the three-dimensional space $XYZ$ when the integration
time $\tau=10^5$. } \label{Fig2}}
\end{figure*}

\begin{figure*}[htbp]
\center{
\includegraphics[scale=0.31]{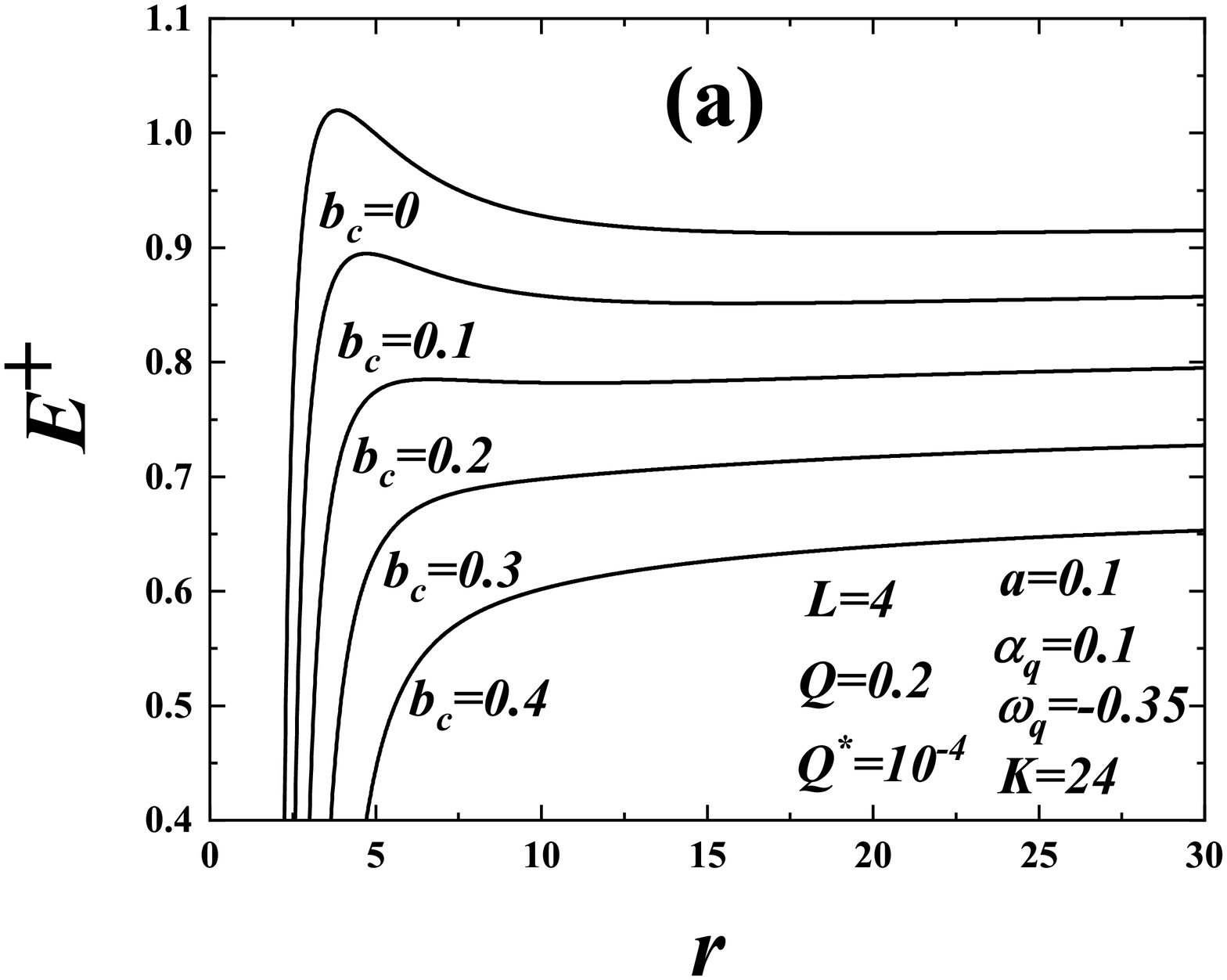}
\includegraphics[scale=0.31]{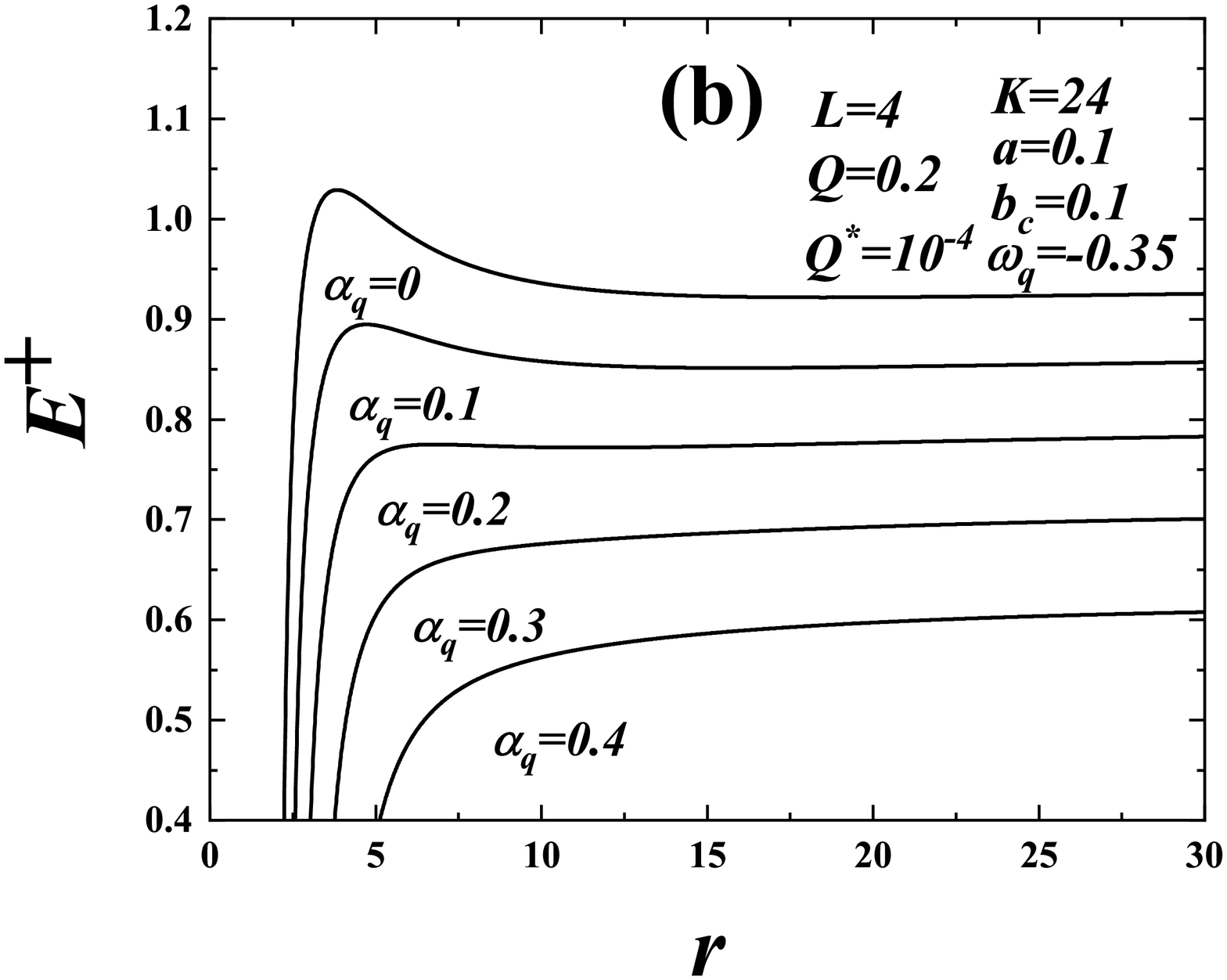}
\includegraphics[scale=0.31]{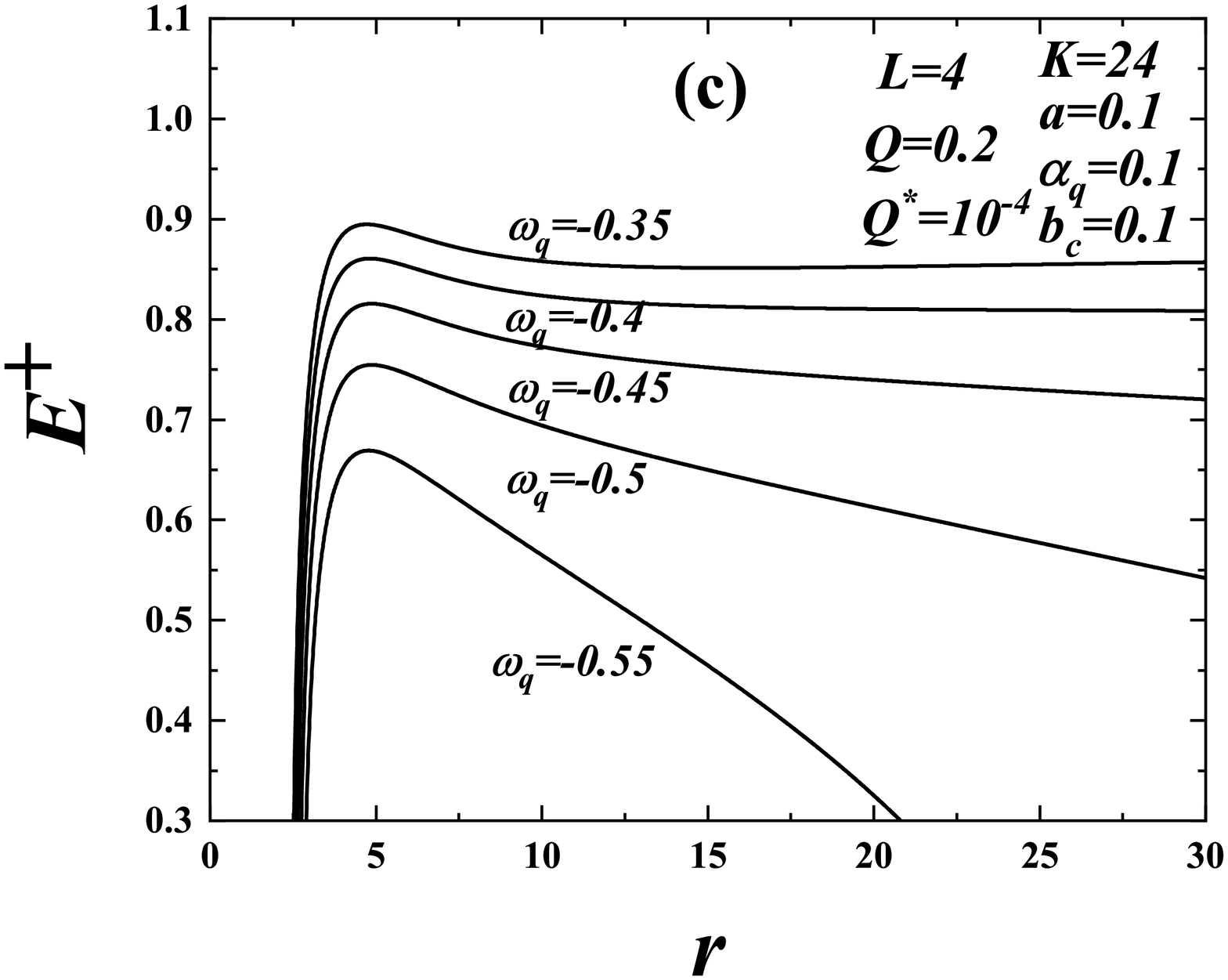}
\includegraphics[scale=0.31]{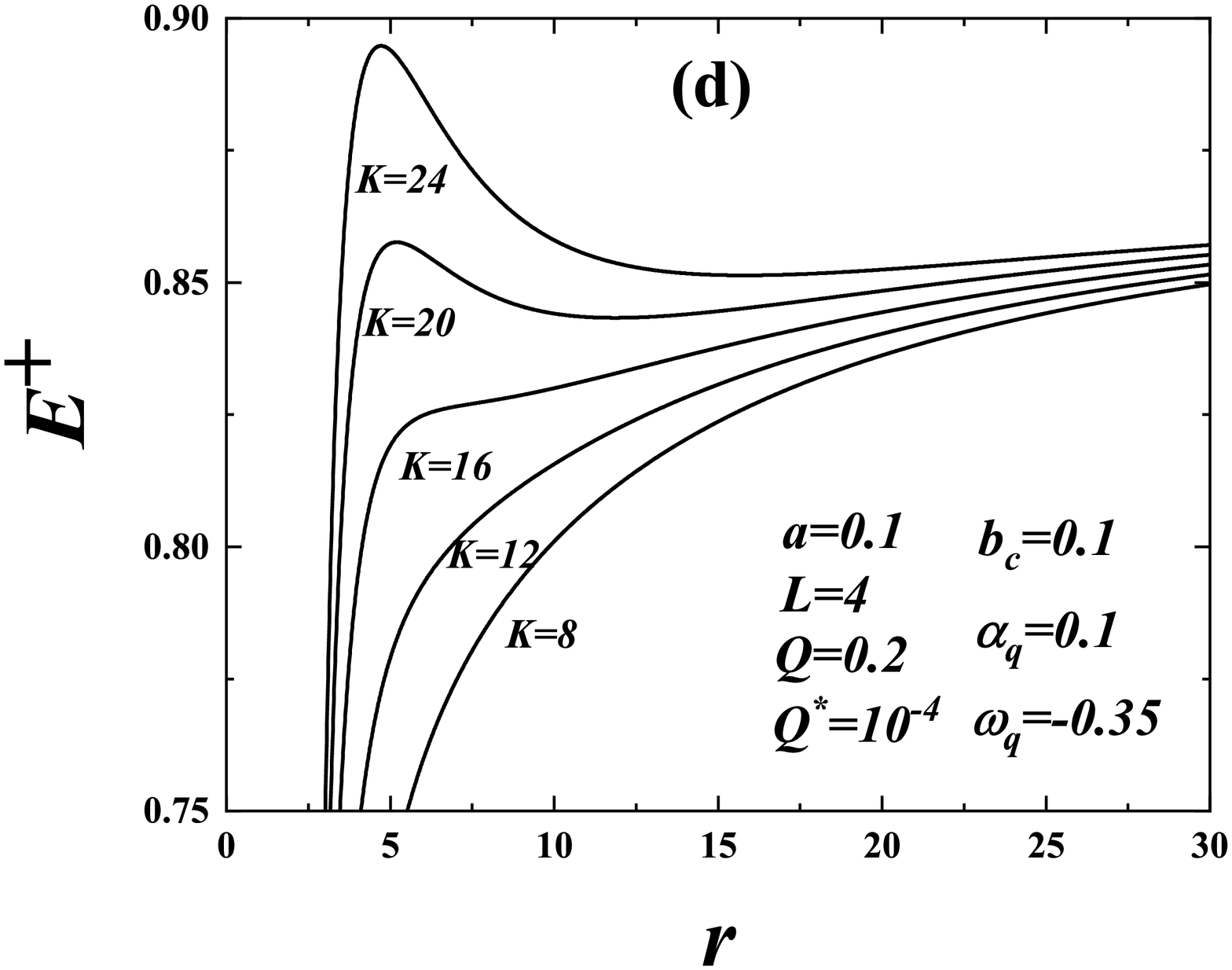}
\caption{Radial effective potentials $E^{+}$ of Eq. (49) in the
three-dimensional configuration. (a): The dependence of $E^{+}$ on
the cloud strings parameter $b_c$ shows the decrease of $E^{+}$
with the increase of $b_c$. (b): The dependence of $E^{+}$ on the
quintessence parameter $\alpha_q$ describes the decrease of
$E^{+}$ with the increase of $\alpha_q$. (c): The dependence of
$E^{+}$ on the quintessential state parameter $\omega_q$ indicates
the increase of $E^{+}$ with the increase of $\omega_q$. (d): The
dependence of $E^{+}$ on the Carter-like constant $K$ exhibits the
increase of $E^{+}$ with the increase of $K$, where $K$ does not
satisfy Eq. (57) with $p_{\theta}=0$ but is freely given and
satisfies Eq. (44) with $p_{\theta}\neq0$. }
 \label{Fig3}}
\end{figure*}

\begin{figure*}[htbp]
\center{
\includegraphics[scale=0.2]{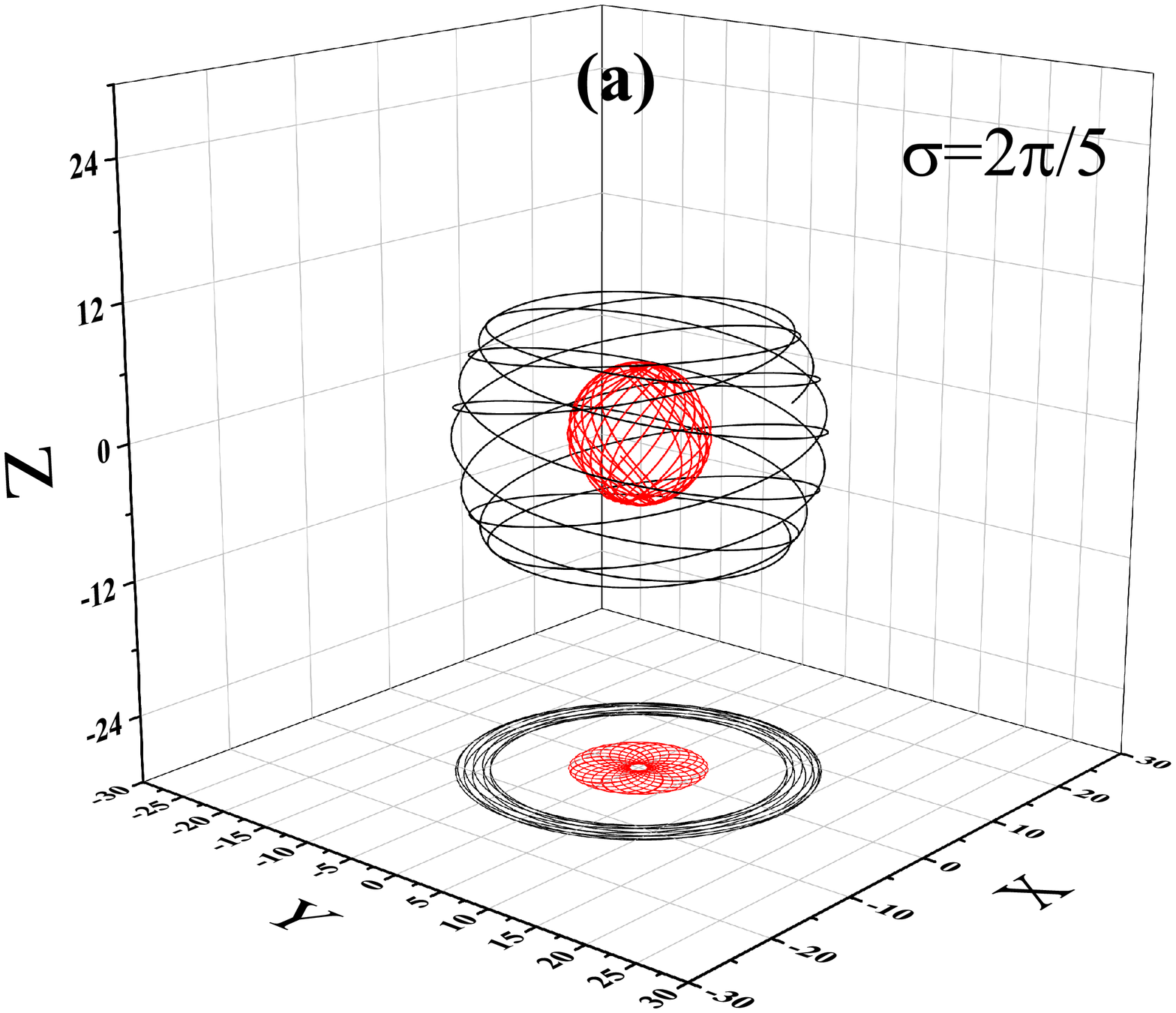}
\includegraphics[scale=0.2]{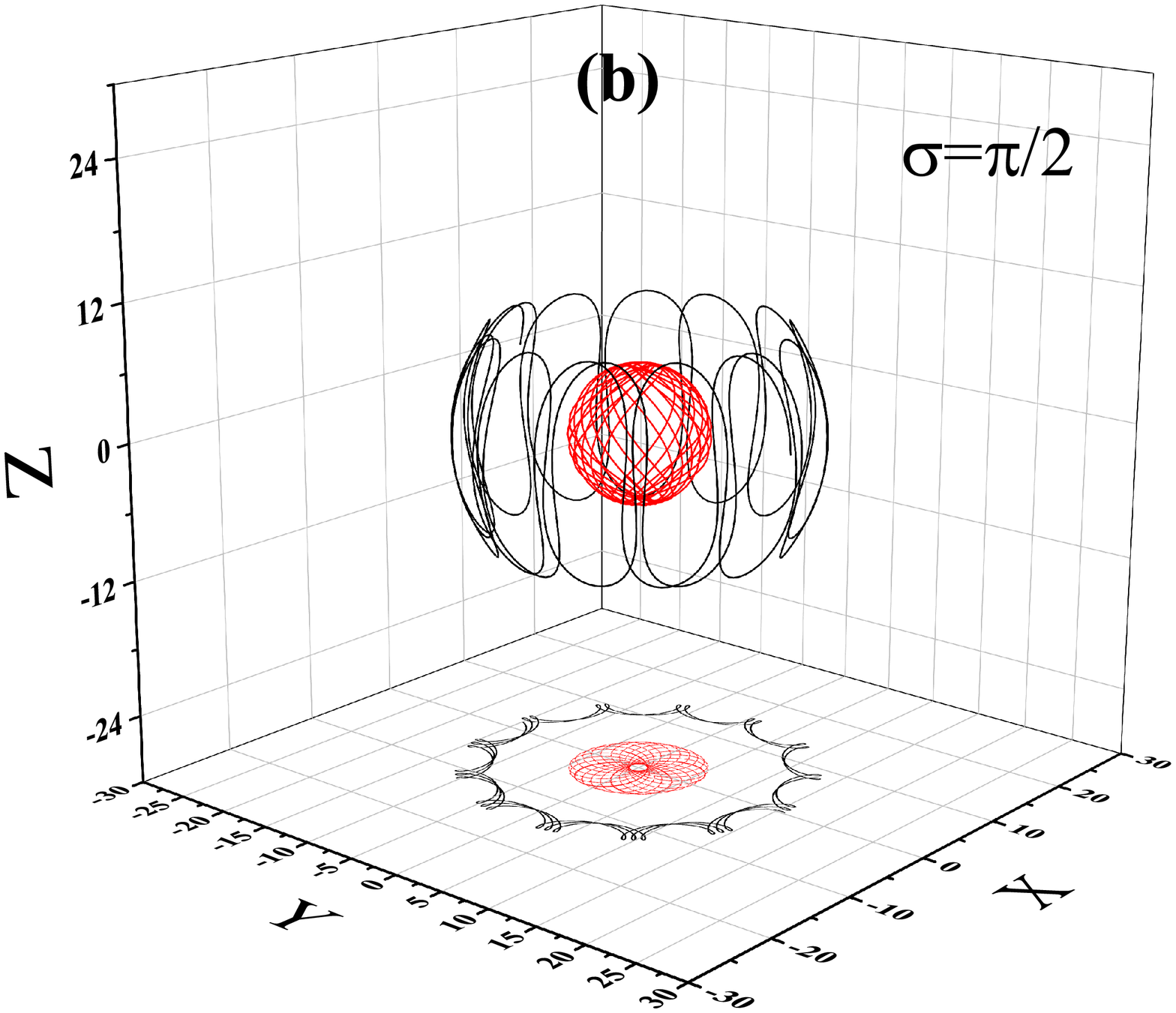}
\includegraphics[scale=0.2]{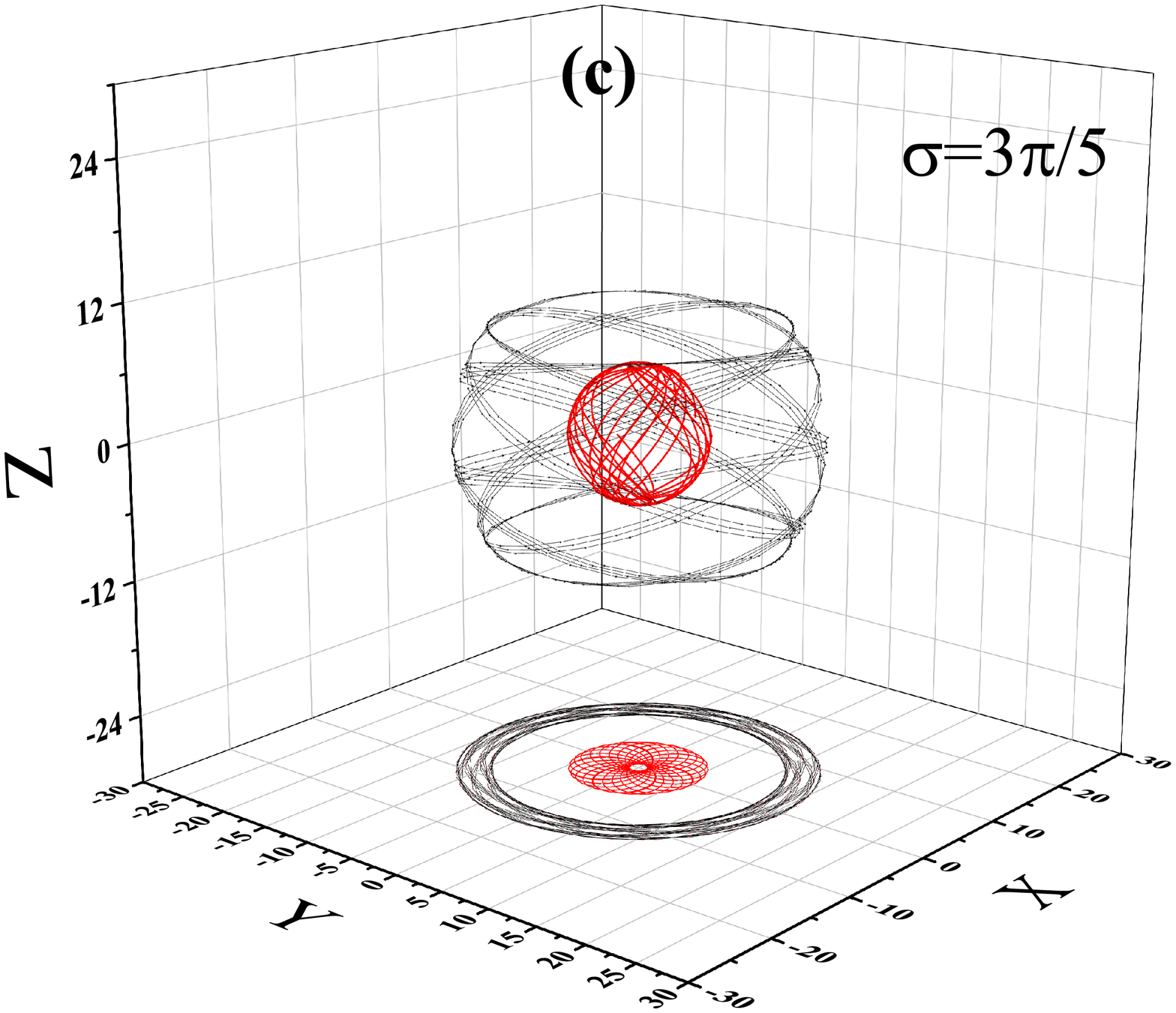}
\caption{Three examples of SSOs and MSSOs in the three-dimensional
space. (a): The SSO colored Black has the initial values
$\sigma=2\pi/5$ and $p_{\theta}=2.64$, and the MSSO colored Red
has the initial values $\sigma=2\pi/5$ and $p_{\theta}=2.46$. (b):
The SSO colored Black has the initial values $\sigma=\pi/2$ and
$p_{\theta}=2.94$, and the MSSO colored Red has the initial values
$\sigma=\pi/2$ and $p_{\theta}=2.71$. (c): The SSO colored Black
has the initial values $\sigma=3\pi/5$ and $p_{\theta}=2.64$, and
the MSSO colored Red has the initial values $\sigma=3\pi/5$ and
$p_{\theta}=2.46$. In these panels, all SSOs have the radii
$R_S=15.79$ and the Carter-like constant $K=24$, and all MSSOs
correspond to the radii $R_M=7.83$, the angular momentum $L=-3.47$
and the Carter-like constant $K=20$. The other parameters are the
same as those of Fig. 3(d). Unlike those in Figs. 1 and 2, the
values of $\sigma$  do not satisfy Eq. (57) with $p_{\theta}=0$
and are no longer constant plane parameters. In fact, the values
of $\sigma$ in the three panels, such as $\sigma=2\pi/5$ in panel
(a), are only the initial values of $\sigma$ but are not invariant
with time. The upper part of each panel corresponds to the
practical trajectories, and the bottom part relates to projections
of the practical trajectories. These orbits still remain spherical
and stable when the integration time $\tau=10^5$.} \label{Fig4}}
\end{figure*}

\begin{figure*}[htbp]
\center{ \vspace{2em}
\includegraphics[scale=0.2]{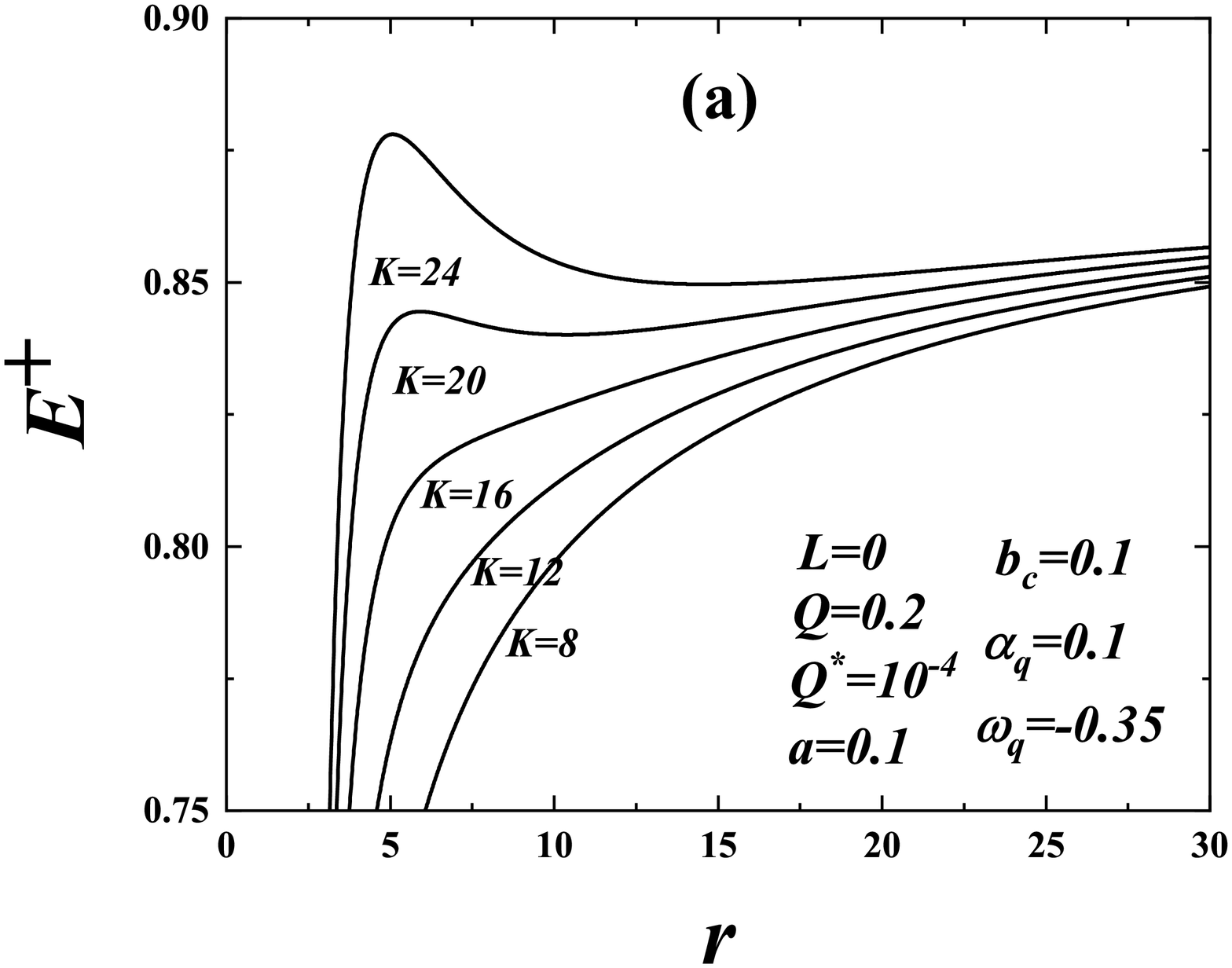}
\includegraphics[scale=0.2]{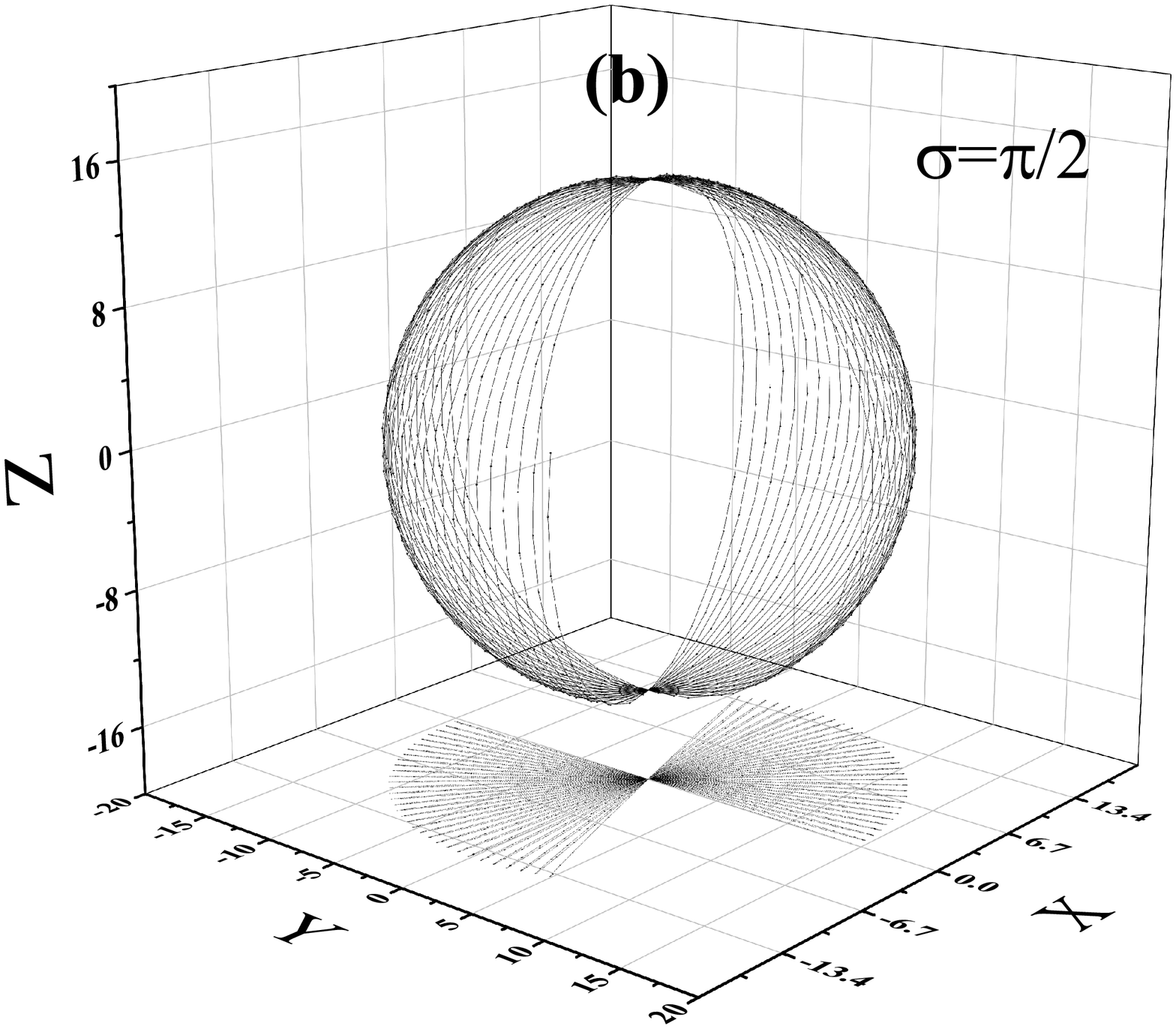}
\includegraphics[scale=0.2]{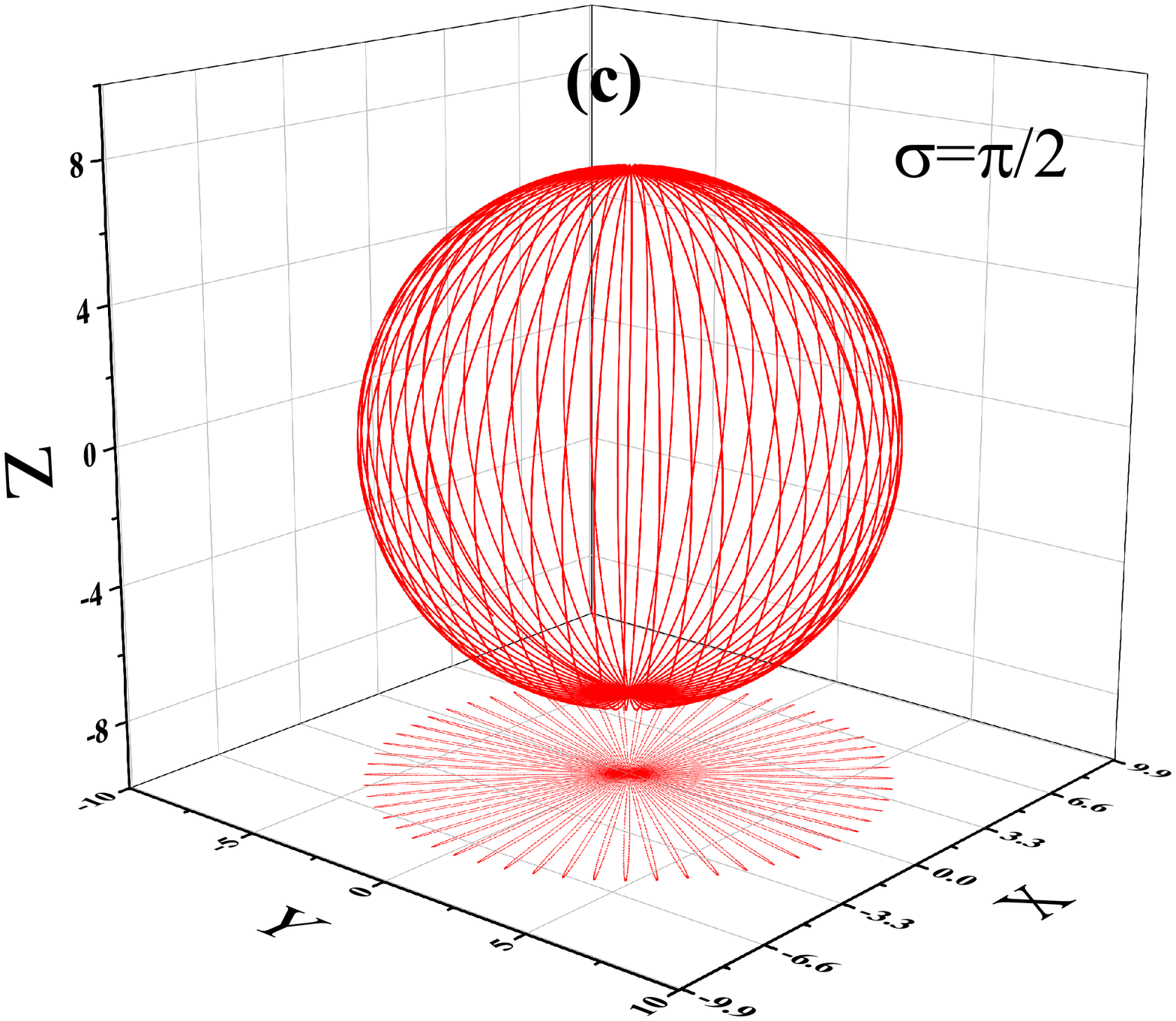}
\caption{(a): Relation between the Carter-like constant $K$ and
radial effective potential $E^{+}$ with zero angular momentum
$L=0$ in the three-dimensional configuration. The potential
increases with $K$ increasing, where $K$ does not satisfy Eq. (57)
with $p_{\theta}=0$ but is freely given and satisfies Eq. (44)
with $p_{\theta}\neq0$. (b): A stable spherical orbit with $K=24$,
the radius $R_S=14.68$ and the initial value $p_{\theta}=4.89$.
(c): A marginally stable spherical orbit with $K=18.54$, the
radius $R_M= 7.53$ and the initial value $p_{\theta}=4.31$. The
other parameters of panels (b) and (c) are those of panel (a).
Unlike those in Figs. 1 and 2, the values of $\sigma$ do not
satisfy Eq. (57) with $p_{\theta}=0$ and are no longer constant
plane parameters. In fact, $\sigma=\pi/2$ is only the initial
value of $\sigma$ but varies with time. The upper part of each
panel corresponds to the practical trajectories, and the bottom
part relates to projections of the practical trajectories. The two
spherical orbits seem to have same sizes but have different radii:
$R_S=14.68$ in  panel (b) and $R_M= 7.53$ in panel (c). These
orbits with vanishing angular momenta for covering whole the range
of the latitudinal coordinate in panels (b) and (c) still remain
spherical and stable when the integration time $\tau=10^5$.
}\label{Fig5}}
\end{figure*}
\end{document}